\documentclass{aa}

\usepackage{graphicx}
\usepackage{txfonts}
\usepackage{bm}

\usepackage{natbib}
\bibpunct{(}{)}{;}{a}{}{,}

\newcommand{\beq}{\begin{equation}}
\newcommand{\eeq}{\end{equation}}
\newcommand{\bea}{\begin{eqnarray}}
\newcommand{\eea}{\end{eqnarray}}
\newcommand{\req}[1]{Eq.~(\ref{#1})}

\newcommand{\aB}{a_\mathrm{B}}
\newcommand{\am}{a_\mathrm{m}}
\newcommand{\dd}{\mathrm{d}}
\newcommand{\gcc}{\mbox{g cm$^{-3}$}}

\newcommand{\Kc}{K_\mathrm{c}}
\newcommand{\Kp}{K_\perp}
\newcommand{\mel}{m_\mathrm{e}}
\newcommand{\mpr}{m_\mathrm{p}}
\newcommand{\mH}{m_\mathrm{H}}
\newcommand{\omc}{\omega_\mathrm{ce}}
\newcommand{\omcp}{\omega_\mathrm{cp}}

\begin{document}

\title{Opacities and spectra of hydrogen atmospheres
      of moderately magnetized neutron stars}
                                                         
\author{
A. Y. Potekhin\inst{1,2,3,4}\thanks{\email{palex@astro.ioffe.ru}}
\and
G. Chabrier\inst{1,5}
\and
W. C. G. Ho\inst{6}
}
\institute{
Centre de Recherche Astrophysique de Lyon, Universit\'e de
Lyon, Universit\'e Lyon 1, Observatoire de Lyon, Ecole Normale
Sup\'erieure de Lyon, CNRS, UMR 5574, 46 all\'ee d'Italie,
69364, Lyon Cedex 07, France
\and
Ioffe Institute,
Politekhnicheskaya 26, St.~Petersburg 194021, Russia
\and
Central Astronomical Observatory at Pulkovo,
Pulkovskoe Shosse 65, 196140 Saint Petersburg, Russia
\and
Isaac Newton Institute of Chile, 
         St.~Petersburg Branch, Russia
\and
School of Physics, University of Exeter, Exeter, UK EX4 4QL
\and
Mathematical Sciences and STAG Research Centre, 
University of Southampton, Southampton
SO17 1BJ, UK
}

\date{Received 16 July 2014 / Accepted 26 September 2014}

%%%%%%%%%%%%%%%%%%%%%%%%%%%%%%%%%%%%%%%%%%%%%%%%%%%%%%%%%
\abstract
{There is observational evidence that central compact
objects (CCOs) in supernova remnants have moderately strong
magnetic fields $B\sim10^{11}$~G. Meanwhile, available
models of partially ionized hydrogen atmospheres of neutron
stars with strong magnetic fields are restricted to 
 $B\gtrsim10^{12}$~G. Extension of the
applicability range of the photosphere models to smaller field
strengths is complicated by a stronger asymmetry of
decentered atomic states and by the importance of excited bound
states.
}{We extend the equation of state and radiative opacities,
presented in previous papers for $10^{12}\mbox{ G}\lesssim B
\lesssim 10^{15}$~G, to weaker fields.
}{
We construct analytical fitting formulae for binding energies,
sizes, and
oscillator strengths for different bound
states of a hydrogen atom moving in moderately
strong magnetic fields and
calculate an extensive database for photoionization cross
sections of such atoms. Using these atomic data, in the
framework of the chemical picture of plasmas we solve
the ionization equilibrium problem and calculate
thermodynamic functions and basic
opacities of partially ionized
hydrogen plasmas at these field strengths. Then plasma
polarizabilities are calculated from the Kramers-Kronig
relation, and the radiative transfer equation for the
coupled normal polarization modes is solved to obtain model
spectra.
}{
An equation of state and radiative opacities 
for a partially ionized hydrogen plasma are obtained
at magnetic fields $B$, temperatures $T$, and densities 
$\rho$ typical 
for atmospheres of CCOs and other isolated neutron stars
with moderately strong magnetic fields.
The first- and second-order thermodynamic functions,
monochromatic radiative opacities, and
Rosseland mean opacities are calculated 
and tabulated, taking account of partial ionization,
for
$3\times10^{10}\mbox{ G}\lesssim B\lesssim 10^{12}$~G,
 $10^5$ K $\lesssim T\lesssim 10^7$ K, 
 and a wide range of densities. Atmosphere models and
spectra are calculated to verify the applicability of the
results and to determine the range of magnetic fields and
effective temperatures where the incomplete ionization of
the hydrogen plasma is important.
}{}


\maketitle

%%%%%%%%%%%%%%%%%%%%%%%%%%%%%%%%%%
\section{Introduction}
\label{sect:intro}

Thermal or thermal-like radiation has been detected from
several classes of neutron stars, which are characterized by
different typical values of magnetic field $B$. Particularly interesting
are isolated neutron stars with clearly observed thermal
emission in quiescence, whose thermal X-ray spectra formed
at the surface are not blended with emission from accreting
matter or magnetosphere \citep[see the list of their
properties in][]{Vigano-ea13}. Most of them have surface
magnetic fields in the range $10^{12}\mbox{ G}\lesssim B
\lesssim 10^{15}$~G, but one class of sources, so called
central compact objects (CCOs) have $B\sim\mbox{a
few}\times(10^{10}$\,--\,$10^{11})$~G
\citep{HalpernGotthelf10,Ho13}. These fields are
sufficiently strong to radically affect properties of
hydrogen atoms and strongly quantize the electrons in the
neutron-star atmosphere, but they are below the field
strengths available in the previously developed models of
strongly magnetized partially ionized hydrogen atmospheres
of neutron stars, which are currently included in the XSPEC
package \citep{XSPEC} under the names NSMAX \citep{HoPC08}
and NSMAXG \citep{Ho14}. Therefore a construction
of neutron-star photosphere models for moderately strong
magnetic fields has become a topical problem. In this paper
we construct such models for
$3\times10^{10}<B<10^{12}$~G. 

We use the theoretical model of a partially ionized hydrogen
plasma (\citealp{PCS99}; hereafter Paper~I) that
was previously used for opacity calculations at
$10^{12}\mbox{~G}\lesssim B \lesssim10^{15}$~G
(\citealp{PC03,PC04}; hereafter Papers~II and~III,
respectively). However, the present task is more arduous,
because the field strength is closer to the atomic unit $B_0
= {\mel^2\,c\,e^3}/{\hbar^3}=2.35\times10^9$~G. Accordingly,
the dimensionless magnetic field parameter $\gamma = B/B_0$
is smaller, and the adiabatic approximation for atomic wave
functions, which is valid at $\gamma\to\infty$, becomes less
adequate, which entails the need to include more terms than
previously in the wave-function expansion beyond this
approximation.

%%%%%%%%%%%%%%%%%%%%%%%%%%%%%%%%%%%%%%%%%%%%%%%%%%%%%%%%%
\begin{figure*}
\centering
\includegraphics[height=.41\textwidth]{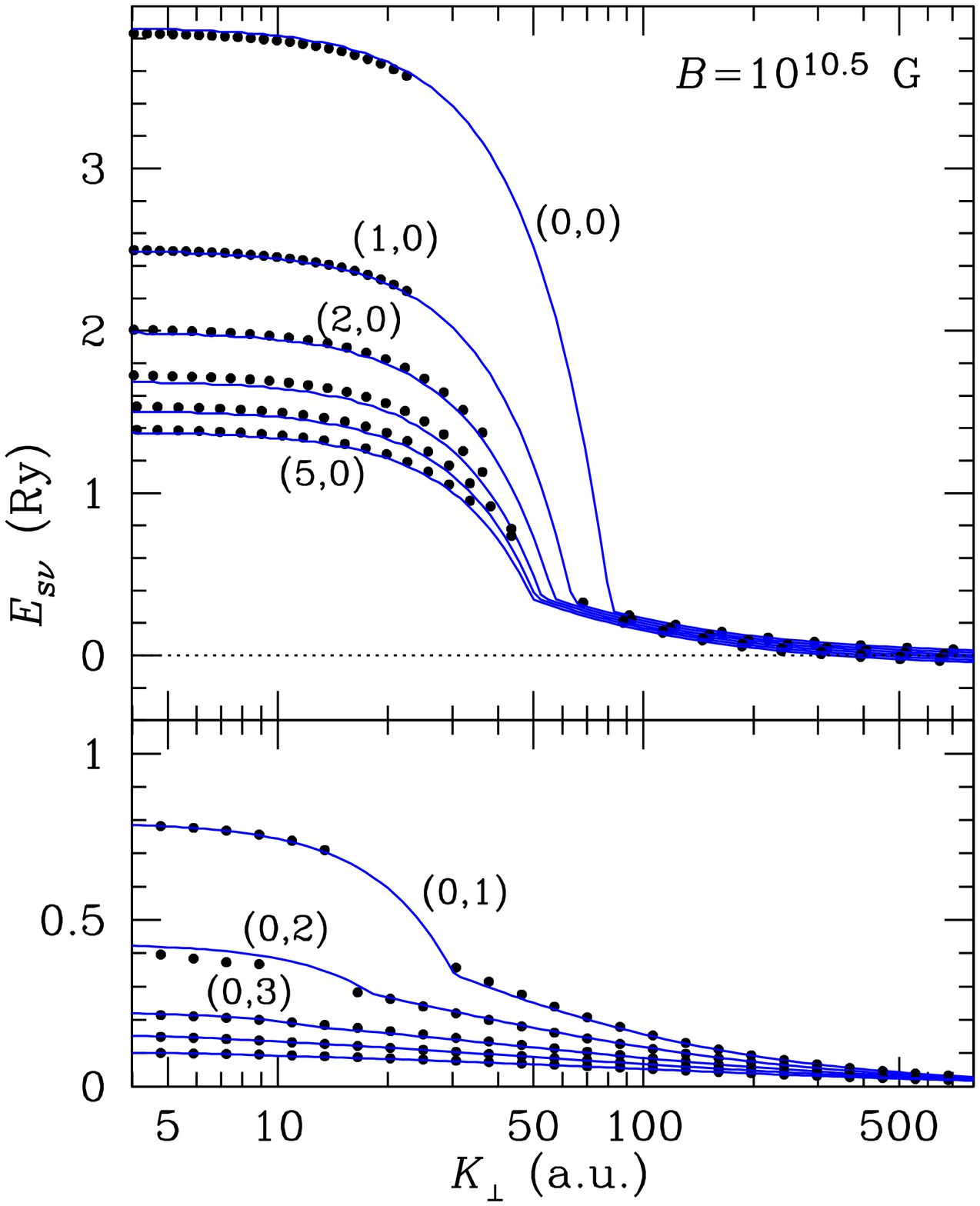}
\includegraphics[height=.41\textwidth]{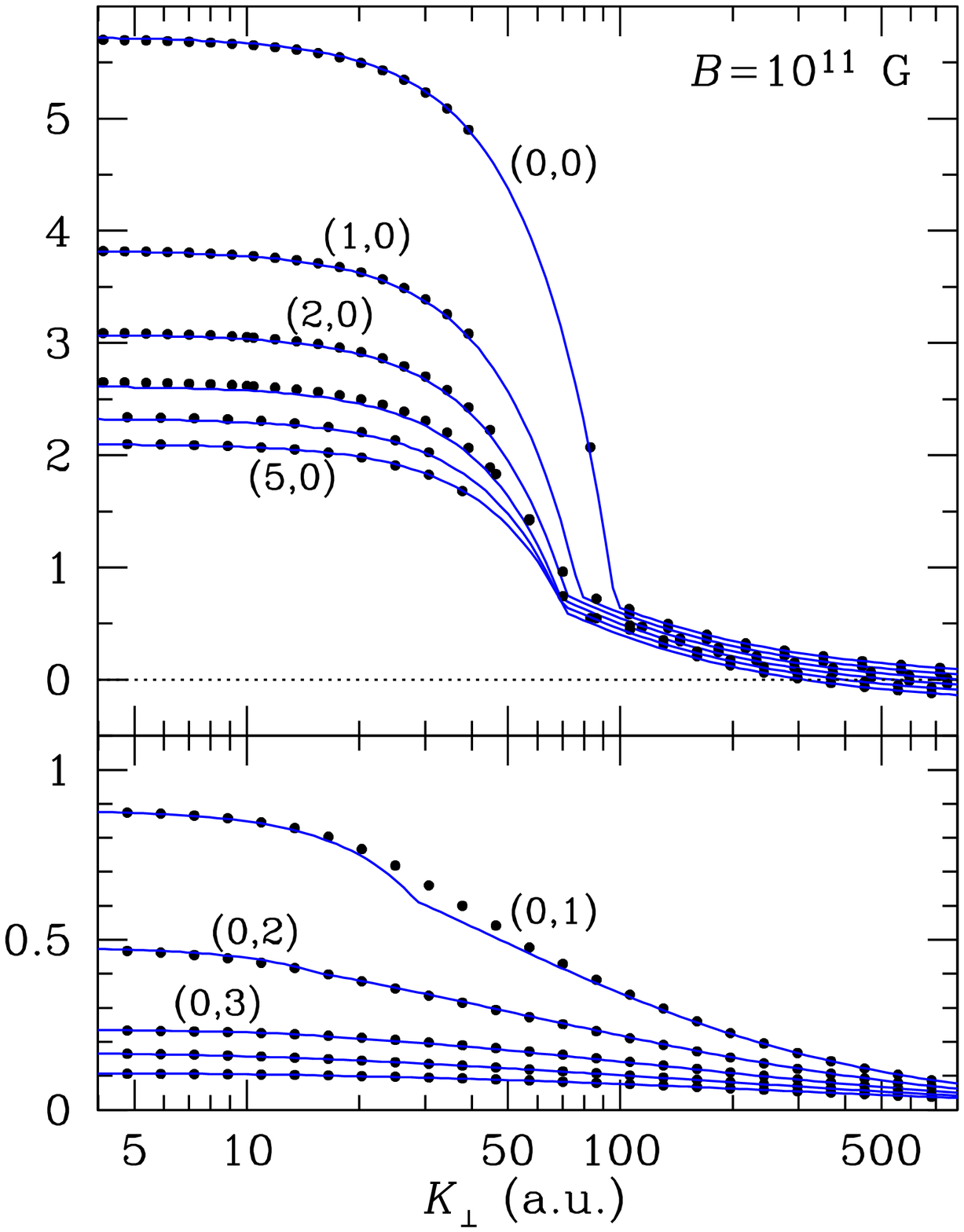}
\includegraphics[height=.41\textwidth]{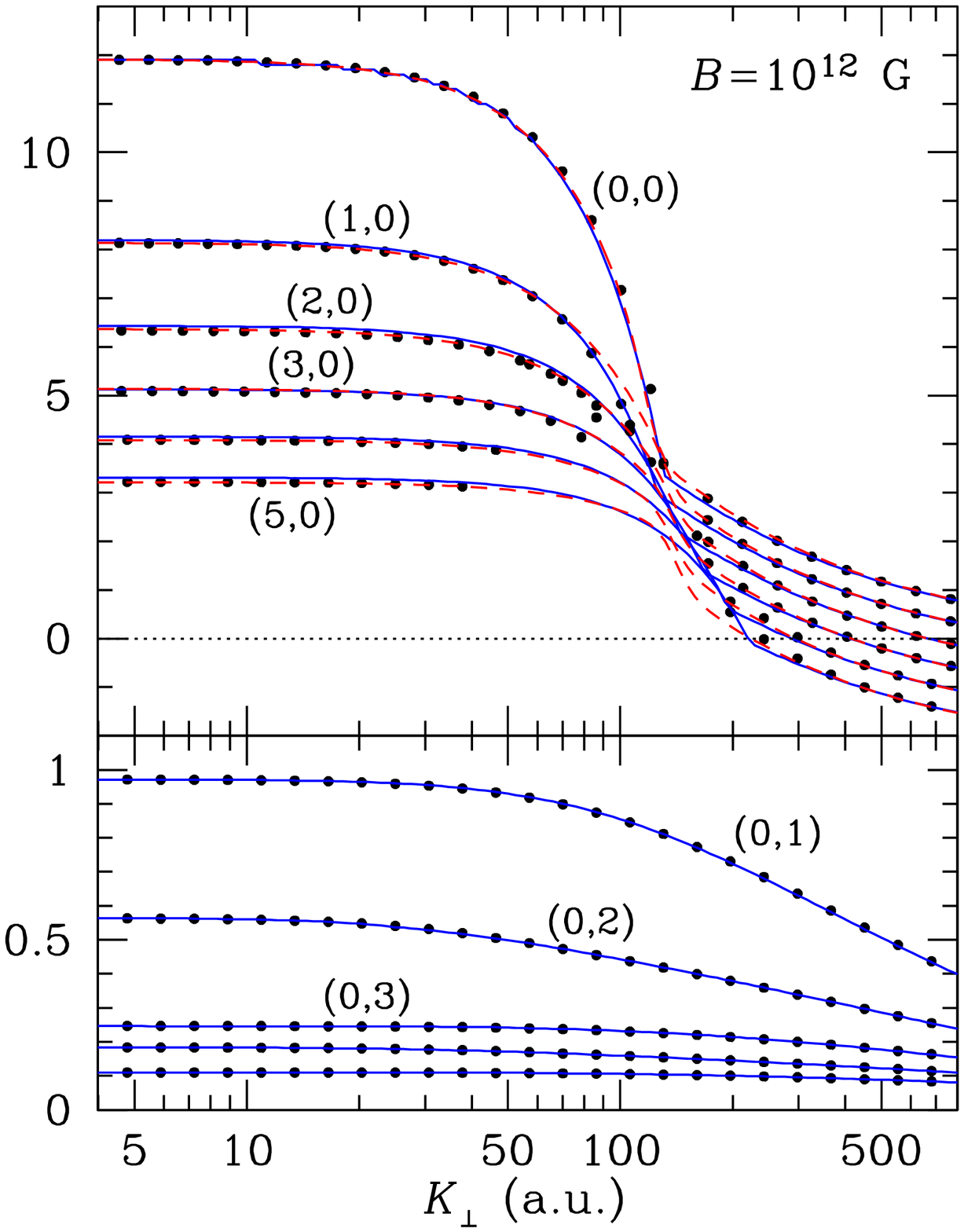}
\caption{Binding energies of different states ($s,\nu$) of a
hydrogen atom as functions of pseudomomentum $\Kp$ at
$B=3.16\times10^{10}$~G (left), $B=10^{11}$~G (middle),
and $B=10^{12}$~G (right). Numerical
calculations (dots) are compared with the analytical
approximation presented in
Appendix~\ref{sect:en} (solid lines). For comparison, the fit
previously developed for $B\gtrsim10^{12}$~G \citep{P98} is
shown by the dashed lines in the upper right panel.
}
\label{fig:enfit14}
\end{figure*}
%%%%%%%%%%%%%%%%%%%%%%%%%%%%%%%%%%%%%%%%%%%%%%%%%%%%%%%%%

In addition, with decreasing $B$, the energy spectrum of the
bound states of a magnetized atom becomes denser, which
necessitates inclusion of more such states in the
consideration. Meanwhile, since $\gamma\gg1$, the center-of
mass motion of the atom noticeably affects the atomic
properties. In order to cope with the problem, we construct
analytical fitting formulae for atomic energies, sizes, and
main oscillator strengths as functions of $B$, discrete
quantum numbers of initial and final states, and 
pseudomomentum $\Kp$, which corresponds to the state of
motion of an atom across the field. These analytical
expressions are valid for $\gamma\sim10$\,--\,1000 and
supplement the previously available fits for larger $\gamma$
\citep{P98}. For bound-free transitions, we calculate
extensive tables of cross sections as functions of $\Kp$ and
photon frequency $\omega$ for a number of bound
states at every given $B$ and interpolate across these
tables to calculate the opacities in the same way as in
Papers~II and~III.

In Sect.~\ref{sect:Hatom} we recall the main properties that
characterize the motion of a hydrogen atom in a magnetic
field. In Sect.~\ref{sect:ioneq} we describe the solution of
the ionization equilibrium problem. Section~\ref{sect:mopa}
contains the summary of the theoretical methods used to
calculate the opacities and polarization vectors of normal
electromagnetic modes in a magnetized plasma. Results of
numerical calculations are presented and discussed in
Sect.~\ref{sect:results}. In Sect:~\ref{sect:concl} we
formulate conclusions. In the Appendix we present analytical
approximations to the results of numerical calculations of
the characteristics of a hydrogen atom moving in a strong
magnetic field (binding energies, quantum-mechanical sizes,
bound-bound transition oscillator strengths) that are used
to solve the ionization equilibrium problem and to calculate
the opacities.

%%%%%%%%%%%%%%%%%%%%%%%%%%%%%%%%%
\section{Hydrogen atom in a strong magnetic field}
\label{sect:Hatom}

Motion of charged particles in a magnetic field $\bm{B}$ is quantized
in discrete Landau levels.
In the nonrelativistic theory, the
energy of the $N$th Landau level equals $N\hbar\omc$
($N=0,1,2,\ldots$), where $\omc=eB/\mel c$ is the
electron cyclotron frequency.
The wave functions
that describe an electron in a magnetic field
have a characteristic transverse scale of the order of the
``magnetic length'' $\am=(\hbar
c/eB)^{1/2}=\aB/\sqrt{\gamma}$, where $\aB$ is the Bohr
radius. 

In a hydrogen atom, the Landau quantization affects motion
of both charged particles, electron and proton. For a
nonmoving atom in a strong magnetic field, there are two
distinct classes of its quantum states: at every value of
the Landau quantum number $N$ and the magnetic quantum
number $-s$ ($N\geq 0$, $s\geq -N$),  there is one tightly
bound state (with ``longitudinal'' quantum number $\nu=0$),
with binding energy growing asymptotically as $(\ln
\gamma)^2$, and an infinite series of loosely-bound states
($\nu=1,2,\ldots$) with binding energies below 1~Ry. The sum
$N+s$ corresponds to the Landau number for the proton. At
$B\gtrsim10^9$~G, the electron-proton binding is possible
only for $N=0$. Therefore we drop $N$ from the bound-state
labeling hereafter. Although the Coulomb interaction mixes
different electron and proton Landau orbitals, this
numbering is unambiguous and convenient at $B\gg B_0$
\citep[see][]{P94}.

The binding energy of a hydrogen atom can be written as
\begin{equation}
 E_{s\nu}
    = E_{s\nu}^\| - \hbar\omcp s, 
\label{E0}
\end{equation}
where so called longitudinal binding energy $E_{s\nu}^\|$ is
positive and corresponds to the relative electron-proton
motion along $\bm{B}$, while the term $-\hbar\omcp s$
diminishes the total binding energy due to transverse
quantum excitations by multiples of the proton cyclotron
energy $\hbar\omcp=\hbar eB/\mpr c=6.305\,(B/10^{12}\mbox{ G})$
eV.
The atom is elongated: its size along the
magnetic field $\bm{B}$ either decreases logarithmically
with increasing $\gamma$
(for  the tightly bound states) or remains nearly constant
(for the loosely bound states),  while the transverse radius
decreases as $\gamma^{-1/2}$.

The astrophysical simulations assume finite temperatures,
hence thermal motion of particles. The theory of motion of a
system of point charges in a constant magnetic field was
reviewed by \citet{JHY83}. The canonical center-of-mass
momentum $\bm{P}$ is not conserved in a magnetic field. A
relevant conserved quantity is pseudomomentum, which for the
H atom equals $\bm{K} = \bm{P} - (e/{2c})\,\bm{B}\times
\bm{r}, $ where $\bm{r}$ connects the proton and the
electron. Early studies of the effects of motion were done 
by \citet{GorkovDzyal,Burkova-ea76,IpatovaMS84}.
\citet{VB88} and \citet{PM93} developed a perturbation
theory for the treatment of atoms moving across the
magnetic field at small transverse pseudomomenta $\Kp$.
Numerical calculations of the energy spectrum of the
hydrogen atom with an accurate treatment of the effects of
motion across a strong magnetic field were performed by
\citet{VDB92} and \citet{P94}. 

At small $\Kp$ the binding energy is
\beq
   E_{s\nu}(\Kp) = E_{s\nu}(0) -
\frac{\Kp^2}{2m_{s\nu}},
\label{E-small-K}
\eeq
where $m_{s\nu}$ is an effective ``transverse mass,''
which is larger than the true atomic mass
$\mH$. When $\Kp$
exceeds some critical value $\Kc$, the atom becomes
\emph{decentered}. Then the electron and proton are
localized near their guiding centers, separated by distance
$r_* = (\aB^2/\hbar)\Kp/\gamma$. At $\Kp\to\infty$,
$E^\|_{s\nu}(\Kp) \sim e^3B/(c\Kp)$. More precisely
\citep{P94},
\beq
  E^\|_{s\nu}(\Kp) 
    = \frac{\mbox{2 Ry}}{\sqrt{\hat{r}_* + (2\nu+1)\hat{r}_*^{3/2}
        + \epsilon_{s\nu}(\hat{r}_*)}}\,,
\label{E-large-K}
\eeq
where $\hat{r}_* \equiv r_*/\aB= (\aB/\hbar)\Kp/\gamma$
and $\epsilon_{s\nu}(\hat{r}_*) \sim O(\hat{r}_*)$.
The transverse
atomic velocity equals $\partial E/\partial \bm{K}$.
Therefore with increasing $\Kp$ the velocity
attains a maximum at
$\Kp=\Kc$ and then decreases,  while the average
electron-proton distance continues to increase. For the
states with $s\neq0$, $E_{s\nu}(\Kp)$ can become
negative due to the term $-\hbar\omcp s$ in \req{E0}.  Such
states are metastable. In essence, they are continuum
resonances. In the transition region at $\Kp\approx\Kc$,
the atomic wave function is a complex superposition of
several orbitals, which describe neighboring states outside
this region. 

The width of the range of $\Kp$ around $\Kc$, where the
decentering proceeds, decreases
with decreasing $B$. At $\gamma\gtrsim\mpr/\mel$ this width
is large, and the transition to the
decentered state is smooth, but at $B\lesssim10^{12}$~G the
width is small compared to $\Kc$, so that a tightly-bound atom
becomes decentered almost abruptly. For this reason, the
previous fitting formulae for the $\Kp$-dependences of
the binding energies and other characteristics of the H atom
were restricted to $\gamma>300$ \citep{P98}. In the Appendix
we present a new set of fitting formulae, applicable at
$10\lesssim\gamma\lesssim10^3$. In the overlap region
$300<\gamma\lesssim10^3$ both sets of fitting expressions
describe the atomic characteristics sufficiently well for
the use in the opacity modeling.

Figure~\ref{fig:enfit14} illustrates the
$\Kp$-dependences of binding energies at
$B=3.16\times10^{10}$~G, $B=10^{11}$~G,
and $10^{12}$~G. The results of numerical calculations,
performed by the method described in \citet{P94}, are
compared with the fitting formulae for 5 lowest tightly
bound and 5 lowest loosely bound quantum states. The gaps in
the series of calculated points for some states are related
to a numerical instability in the transition region around
$\Kc$, where no single Landau orbital is clearly leading and
the energy levels experience anticrossings \citep[see
discussion in][]{P94}. In these cases the analytical fits
are more reliable for calculations of the ionization
equilibrium and opacities, which involve integrals over
$\Kp$ (see below). In the case of tightly bound states
at $B=10^{12}$~G, the previous fit for $\gamma>300$
\citep{P98} is also shown. Appreciable differences between
the two fits are observed only 
in the transition region $\Kp\sim\Kc$, where the
anticrossings occur. For the loosely-bound states, the two
fits nearly coincide at this field strength.

%%%%%%%%%%%%%%%%%%%%%%%%%%%%%%%%%
\section{Ionization equilibrium and equation of state}
\label{sect:ioneq}

For photosphere simulations, it is necessary to determine
the fractions of different bound states, which affect
the spectral features via bound-bound and
bound-free absorptions. Solution to this problem is
laborious and ambiguous. The principal difficulty in the
chemical picture of plasmas is the necessity to distinguish
the bound and free electrons and ``attribute'' the bound
electrons to certain nuclei \citep[see, e.g.,][and
references therein]{Rogers00}. Current
approaches to the solution of this problem are based, as a
rule, on the concept of so called occupation probabilities
of quantum states. In the case of strong magnetic fields, 
the occupation probabilities depend not only on the discrete
quantum numbers, but also on the transverse pseudomomentum
$\Kp$.

The momentum projections on the magnetic field have the
usual Maxwellian distributions at thermodynamic equilibrium
for all plasma particles. For transverse motion, however, we
have the discrete Boltzmann distribution over Landau numbers
for electrons and protons, whereas the
transverse momenta of H atoms in a state $(s,\nu)$ have a
distribution $p_{s\nu}(\Kp)$, which is not known in
advance. We adhere to the definition of
$p_{s\nu}(\Kp)$ in Paper~I, so that
$
  2\pi\int_0^\infty p_{s\nu}(\Kp)\,\Kp\,\dd \Kp
            =1.
$
Ionization equilibrium is given by minimization of
the Helmholtz free energy $F$ with respect to particle
numbers, keeping volume $V$ and the total number density of protons
(free and bound)
$
  n_0 = \rho/\mH
$
constant, and the number of electrons equal to that of
protons because of the overall electrical neutrality.
The free energy is written as
\beq
   F = F_\mathrm{id}^\mathrm{e} + F_\mathrm{id}^\mathrm{p} 
       + F_\mathrm{ex} + F_\mathrm{at},
\label{Fren}
\eeq
where $F_\mathrm{id}^\mathrm{e}$,
$F_\mathrm{id}^\mathrm{p}$
are the free energies of ideal gases of the electrons and
protons, respectively, and
$F_\mathrm{ex}$ takes into account the Coulomb
plasma nonideality and the
nonideal contribution which arises from interactions of
bound species with each other and with the electrons and
protons. Finally, $F_\mathrm{at}$ is the contribution of the
atomic gas, including the kinetic and internal degrees of
freedom. The formulae for each term in \req{Fren} are given in
Papers~I and II. In particular,
\beq
 F_\mathrm{at} =
     T V \sum_{s\nu} n_{s\nu}
     \int_0^\infty\!\!\!\ln\left[n_{s\nu} \lambda_\mathrm{H}^3 
        \frac{w_{s\nu}(\Kp) }{ \exp(1) \mathcal{Z}_{s\nu}}\right]
       p_{s\nu}(\Kp)2\pi \Kp\dd \Kp,
\eeq
where $n_{s\nu}$ is the number density of the H atoms with
given discrete numbers $s$ and $\nu$ (any $\Kp$),
$w_{s\nu}(\Kp)$ are the occupation probabilities,
$\lambda_\mathrm{H} = [{2\pi\hbar^2}/({ \mH T})]^{1/2}$
is the thermal wavelength of the atom, and
\begin{equation}
   \mathcal{Z}_{s\nu} =
\frac{ 1 }{ \mH T } 
          \int_0^\infty w_{s\nu}(\Kp)\,
           \mathrm{e}^{E_{s\nu}(\Kp)/ T}
            \Kp \dd \Kp
\label{Z-int}
\end{equation}
is the partition function, which includes the continuous
distribution over $\Kp$.
In all mathematical expressions, temperature $T$ is in
energy units. As in Paper~I, we supplement \req{Fren}
by additional terms due to the molecules H$_n$
($n\geq2$) using approximate
formulae for the characteristics of H$_n$ from
\citet{Lai01}. Since the latter do not take full
account of the motion effects, the results are reliable only
when the abundance of H$_n$ is small, which restricts our
treatment to $T\gtrsim10^5$~K.

%%%%%%%%%%%%%%%%%%%%%%%%%%%%%%%%%%%%%%%%%%%%%%%%%%%%%%%%%
\begin{figure*}
\centering
\includegraphics[height=.43\textwidth]{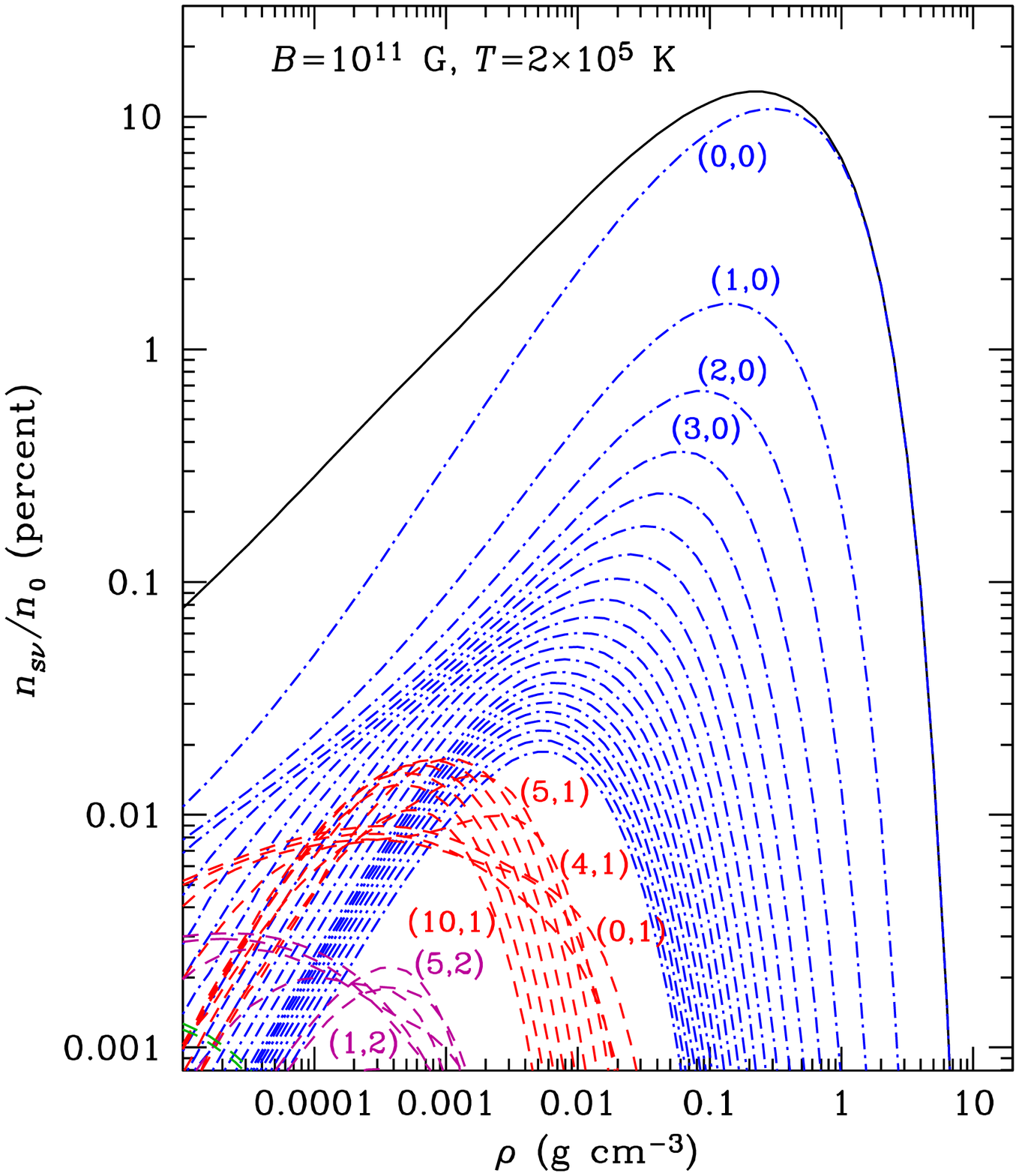}
\includegraphics[height=.43\textwidth]{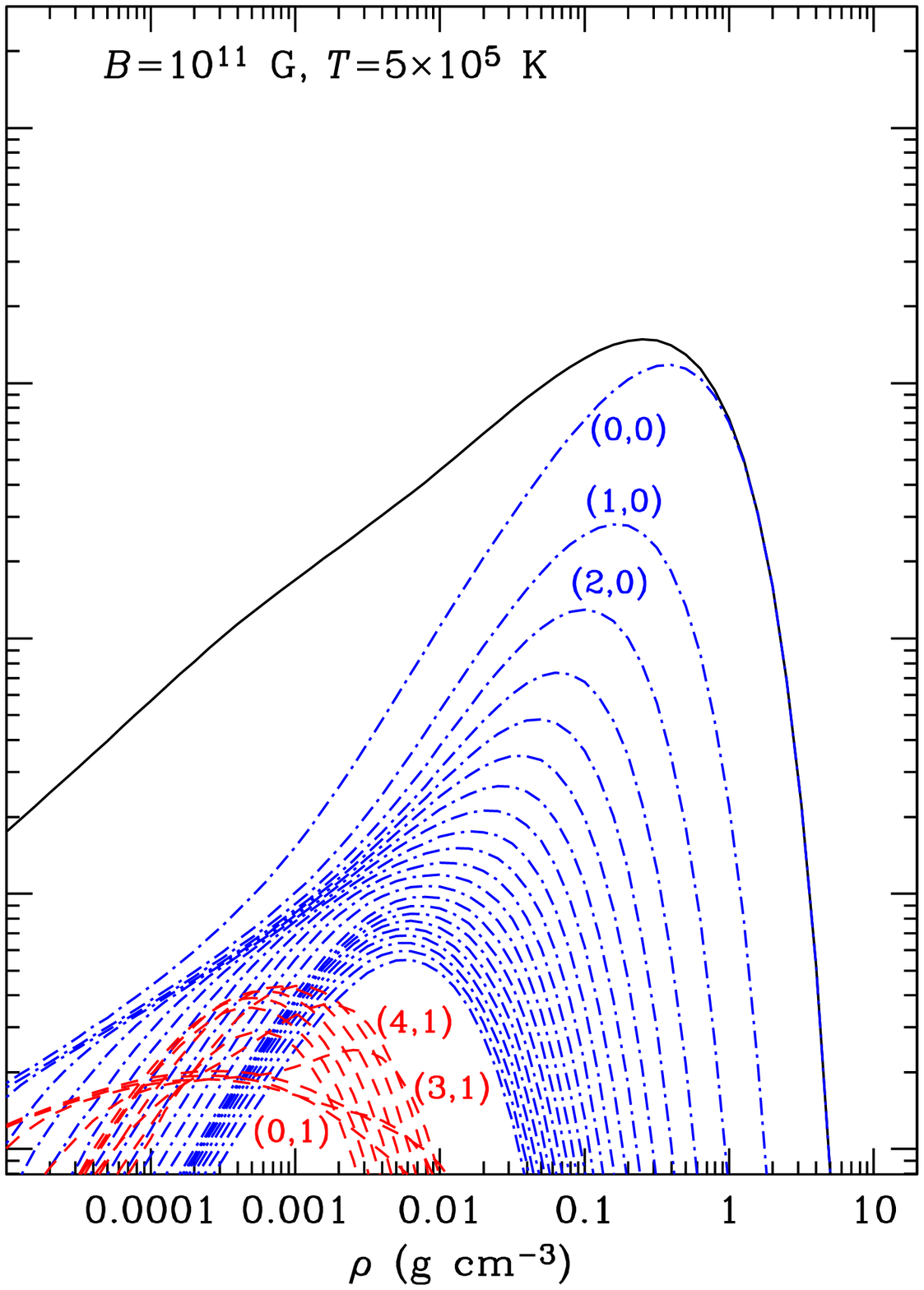}
\includegraphics[height=.43\textwidth]{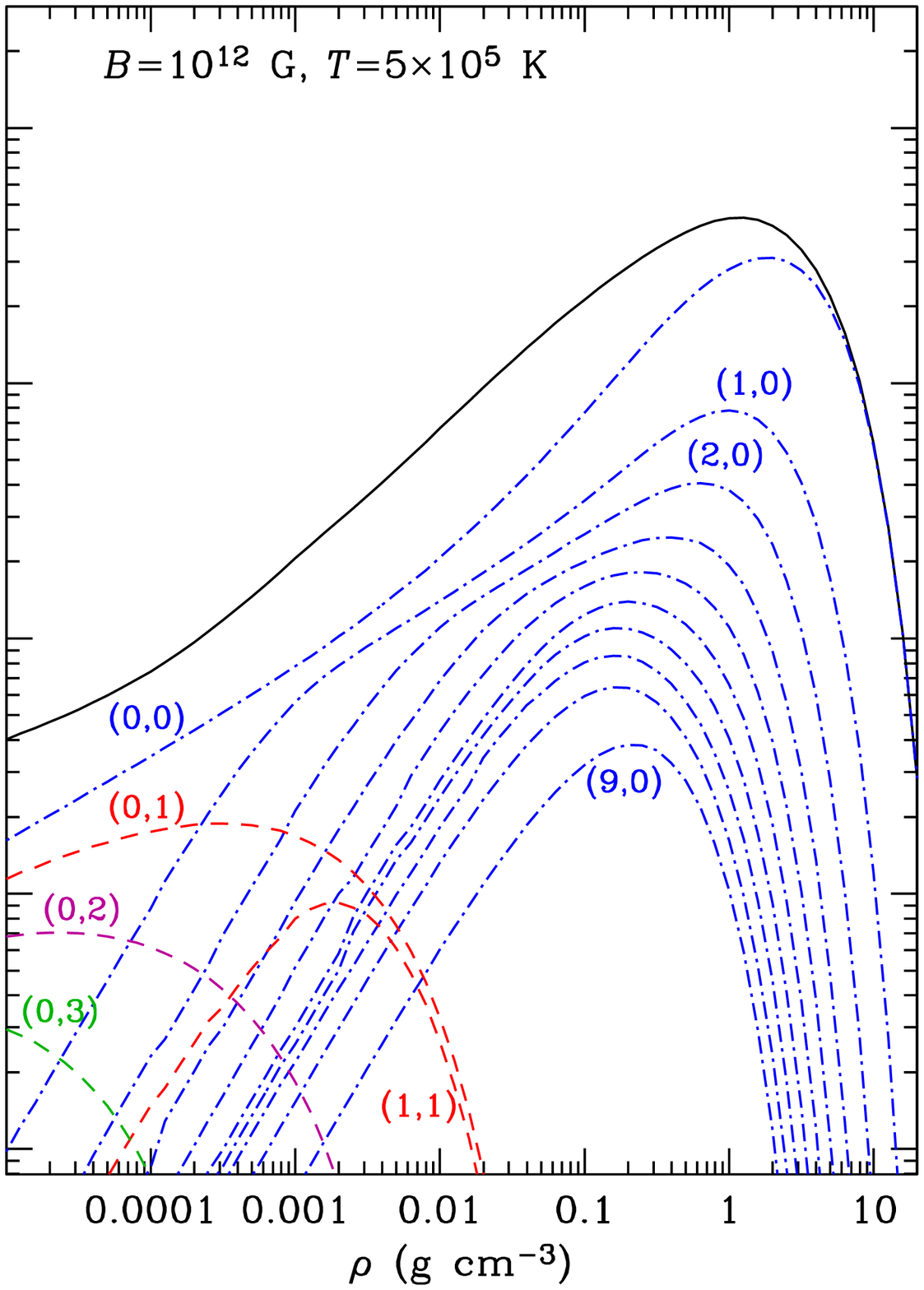}
\caption{Fractional abundances of different
nondissolved bound states $(s,\nu)$ with respect to the
total number of electrons (free and bound), $n_{s\nu}/n_0$,
as functions of mass density $\rho$, for $B=10^{11}$~G and
$T=2\times10^5$~K (left panel), $B=10^{11}$~G and
$T=5\times10^5$~K (middle panel), and $B=10^{12}$~G and
$T=5\times10^5$~K (right panel). The fractional abundances
of tightly bound states ($\nu=0$) are plotted by dot-dashed lines,
and loosely bound states ($\nu\geq1$) by dashed lines.
The solid lines show the total
fractional abundance of the atoms that contribute to
bound-bound and bound-free opacities.
}
\label{fig:ie}
\end{figure*}
%%%%%%%%%%%%%%%%%%%%%%%%%%%%%%%%%%%%%%%%%%%%%%%%%%%%%%%%%

Once the free energy is obtained, its derivatives  over
$\rho$ and $T$ and their combinations provide the other
thermodynamic functions. However, the atomic partial number
fractions $x_{s\nu}=n_{s\nu}/n_0$ that are evaluated in the
course of the free energy minimization cannot be used
directly to calculate opacities. At the considered
conditions, interactions between different plasma particles
give rise to a significant fraction of clusters. Such
clusters contribute to the equation of state similarly to the atoms,
lowering the pressure, but their radiation-absorption
properties differ from  those of an isolated
atom. Therefore we should not count them in the fraction of
atoms $x_\mathrm{H}$ that contribute to the bound-bound and
bound-free opacities. Analogously, at low $\rho$ we should
not include in $x_\mathrm{H}$ the highly excited states that
do not satisfy the  \citet{InglisTeller} criterion of
spectral line  merging, being strongly perturbed by plasma
microfields. Such states form the so called optical
pseudo-continuum  \citep[e.g.,][]{DappenAM87}.  Thus, we
discriminate between the atoms that keep their identity and
the ``dissolved'' states that are strongly perturbed by the
plasma environment. This distinction between the
``thermodynamic'' and ``optical'' neutral fractions is
inevitable in the chemical picture of a plasma 
\citep[see, e.g.,][for discussion]{P96b}. The
fraction of truly bound atoms is evaluated with the use of
the occupation probability formalism. At every $s$, $\nu$,
and $\Kp$, we calculate  the ``optical'' occupation
probability  $w_{s\nu}^\mathrm{o}(\Kp)$, replacing the
Inglis-Teller criterion by an approximate criterion based on
the average atomic  size [Eq.~(14) of \citet{PP95}]. The
fraction of weakly perturbed atoms, which contribute to the
bound-bound and  bound-free opacities, constitutes a
fraction $w_{\nu s}^\mathrm{o}(\Kp)/w_{\nu s}(\Kp) < 1$  of
the total number of atoms. Here, $w_{\nu s}(\Kp)$ is the
``thermodynamic'' occupation probability derived from  the
free energy, which enters the generalized partition function
(\ref{Z-int}). 

Figure \ref{fig:ie} illustrates the dependences of the
fractional abundances of different weakly perturbed atomic
states on $\rho$, $T$, and $B$. The left panel shows the
case of $B=10^{11}$~G and relatively low temperature
$T=2\times10^5$~K. With increasing density, abundance of the
bound states first increases at low $\rho$ according to the
Saha equation and then decreases at higher $\rho$ because of
pressure ionization. The middle panel demonstrates the case
of a higher temperature $T=5\times10^5$~K. In this case, all
bound states are less abundant because of the thermal
ionization. The right panel demonstrates the case of
$T=5\times10^5$~K and a stronger field, $B=10^{12}$~G. Here,
the abundance of the atoms increases compared to the middle
panel because of the larger binding energies. Pressure
ionization starts at larger $\rho$, because the atomic sizes
decrease with increasing $B$. The number of tightly bound
states on the right panel is limited to ten ($0\leq
s\leq9$), the number of states with $\nu=1$ to two
($s=0,1$), and all states with $\nu\geq2$ have $s=0$,
because the states with nonzero $s$ and $\nu\geq2$ are
merged into continuum at $B=10^{12}$~G.

%%%%%%%%%%%%%%%%%%%%%%%%%%%%%%%%%
\section{Polarization modes and opacities}
\label{sect:mopa}

%--------------------------------------------------------
\subsection{Polarization modes}
\label{sect:modes}

Propagation of radiation in magnetized plasmas
was discussed in many 
papers and monographs (e.g., \citealp{Ginzburg}).
In coordinates with the $z$-axis along
$\bm{B}$, the plasma dielectric tensor is
\citep{Ginzburg}
\begin{equation}
  \bm{\varepsilon} = \mathbf{I} + 4\pi\bm{\chi}
          = 
 \left( \begin{array}{ccc}
 \varepsilon_\perp & \mathrm{i} \varepsilon_\wedge & 0 \\
 -\mathrm{i}\varepsilon_\wedge & \varepsilon_\perp & 0 \\
 0 & 0 & \varepsilon_\| 
 \end{array} \right),
\label{eps-p}
\end{equation}
where $\mathbf{I}$ is the unit tensor, 
$\bm{\chi}=\bm{\chi}^\mathrm{H}+\mathrm{i}\bm{\chi}^\mathrm{A}$
is the complex polarizability tensor of plasma,
$\bm{\chi}^\mathrm{H}$ and $\bm{\chi}^\mathrm{A}$ are its
Hermitian and anti-Hermitian parts, respectively.  This
tensor becomes diagonal, 
$\bm{\chi}=\textrm{diag}(\chi_{+1},\chi_{-1},\chi_0)$, in
the cyclic (or rotating) coordinates with unit vectors
$\hat{\mathbf{e}}_{\pm1}=(\hat{\mathbf{e}}_{x} \pm
\mathrm{i}\hat{\mathbf{e}}_{y})/\sqrt{2}$,
$\hat{\mathbf{e}}_{0}=\hat{\mathbf{e}}_{z}$.

At photon energies $\hbar\omega$ much higher than
\beq
   \hbar\omega_\mathrm{pl} =\left(\frac{4\pi\hbar^2 e^2 n_\mathrm{e} 
                      }{ \mel} \right)^{1/2}
                    = 28.7\,\rho^{1/2}  \mathrm{~eV},
\label{plafreqe}
\eeq
where  $\omega_\mathrm{pl}$ is the electron plasma
frequency and $\rho$ is in \gcc,
radiation propagates in the form of extraordinary
(hereafter labeled by $j=1$) and ordinary
($j=2$) normal modes. These modes have different
polarization vectors $\bm{e}_j$ and different absorption and
scattering coefficients, which depend on the angle
$\theta_B$ between the photon wave vector $\bm{k}$ and $\bm{B}$
(e.g., \citealp{KaminkerPS82}). The two modes interact with each
other via scattering. 
\citet{GP73} formulated the radiative transfer
problem in terms of these modes.

At a fixed photon frequency $\omega$,
the absorption opacity $\kappa_j^\mathrm{a}(\theta_B)$ 
in each mode $j$ and scattering opacities
$\kappa_{jj'}^\mathrm{s}(\theta_B)$ 
from mode $j$ into mode $j'$ can be
presented as (e.g., \citealp{KaminkerPS82})
\bea
&&\hspace*{-.7em}
   \kappa_j^\mathrm{a}(\theta_B) = \frac{1}{\mH} \sum_{\alpha=-1}^1
     |e_{j,\alpha}(\theta_B)|^2 \,\sigma_\alpha^\mathrm{a},
\label{kappa-a}
\\&&\hspace*{-.7em}
 \kappa_{jj'}^\mathrm{s}(\theta_B) \!=\!\!
%     \left[ 
     {\frac34}
\!\!
  \sum_{\alpha=-1}^1 \!\!
     |e_{j,\alpha}(\theta_B)|^2 \,
%  \nonumber\\&&\times
     \frac{\sigma_\alpha^\mathrm{s}}{ \mH}\int_0^\pi \!\!\!
       |e_{j',\alpha}(\theta_B')|^2\sin\theta_B'\,\mathrm{d}\theta_B',
%     \right],
\label{kappa-s}
\eea
where $\alpha=0,\pm1$,
 $e_{j,0}=e_{j,z}$ and
$e_{j,\pm1}=(e_{j,x}\pm \mathrm{i} e_{j,y})/\sqrt{2}$
are the components of $\bm{e}_j$ in the cyclic coordinates.
The cross sections $\sigma_\alpha$ depend on 
$\omega$ and $\alpha$, but not on $j$ or $\theta_B$.
The total scattering opacity from mode $j$ is
$\kappa_j^\mathrm{s}=\kappa_{j1}^\mathrm{s}+\kappa_{j2}^\mathrm{s}$,
and the total extinction opacity is
$\kappa_j=\kappa_j^\mathrm{a}+\kappa_j^\mathrm{s}$.

%--------------------------------------------------------
\subsection{Scattering}

Scattering cross-sections in neutron-star photospheres are
well known
\citep{Ventura79,KaminkerPS82,Mesz}.
For $\alpha=-1$, the photon-electron scattering has a
resonance at $\omc$. Outside a
narrow (about the Doppler width) frequency interval around
$\omc$, the cross sections for the basic polarizations
$\alpha=0,\pm1$ are
\begin{equation}
    \sigma_\alpha^\mathrm{s,e} =
          \frac{\omega^2}{(\omega+\alpha\omc)^2
             +\nu_{\mathrm{e},\alpha}^2}\, \sigma_\mathrm{T},
\label{sigma-se}
\end{equation}
where $\sigma_\mathrm{T}=({8\pi}/{3})({e^2}/{\mel
        c^2})^2$ is the nonmagnetic Thomson cross
section,
and
$\nu_{\mathrm{e},\alpha}$ are effective damping factors
(see below).

The photon-ion scattering cross section looks
analogously,
\begin{equation}
    \sigma_\alpha^\mathrm{s,i} =
      \left( \frac{\mel}{\mpr}\right)^2
          \frac{\omega^2}{(\omega-\alpha\omcp)^2
             +\nu_{\mathrm{i},\alpha}^2}\, \sigma_\mathrm{T}.
\label{sigma-sp}
\end{equation}
The resonance at $\omcp$ due to the
scattering on ions can be important in superstrong fields.

In each case, the damping factor $\nu_{\mathrm{e},\alpha}$
or $\nu_{\mathrm{i},\alpha}$
is equal to the half of the total
rate of spontaneous and collisional decay of the 
state with energy $\hbar\omega$ (see discussion in \citealp{PL07}). 
The  spontaneous decay rates are
\begin{equation}
    2\nu_{\mathrm{e}}^\mathrm{s} =
        \frac43\,\frac{e^2}{\mel c^3}\,\omega^2,
\quad
    2\nu_{\mathrm{i}}^\mathrm{s} =
        \frac43\,\frac{e^2}{\mpr c^3}\,\omega^2.
\end{equation}
For the proton-electron plasmas,
the damping factors that include the scattering and 
free-free processes can be approximately written as
(Paper~II)
\begin{equation}
   \nu_{\mathrm{e},\alpha} = \nu_{\mathrm{e}}^\mathrm{s}
        + \nu_{\alpha}^{\mathrm{ff}}(\omc),
\quad
  \nu_{\mathrm{e},\alpha} = \nu_{\mathrm{e}}^\mathrm{s}
        + (\mel/\mpr)\nu_{\alpha}^{\mathrm{ff}}(\omcp),
\end{equation}
where $\nu_{\alpha}^{\mathrm{ff}}(\omega)$ is the effective
free-free frequency given by \req{nu-ff} below.

%--------------------------------------------------------
\subsection{Absorption}

%--------------------------------------------------------
\subsubsection{Cyclotron absorption}

Without magnetic field, absorption of a photon by a free
electron is impossible without involvement of a third
particle, which would accept the difference between the
values of the total momentum of the electron and the photon
before and after the absorption. In a quantizing magnetic
field, a photon can be absorbed or emitted by a free
electron in a transition between Landau levels. In the
nonrelativistic or dipole approximation, such transitions
occur between the neighboring levels at the frequency
$\omc$. In the relativistic theory, the multipole expansion
leads to an appearance of cyclotron harmonics
\citep{Zheleznyakov}. Absorption cross-sections at these
harmonics were derived in the Born approximation by
\citet{PavlovSY80} and represented in a compact form by
\citet{SuleimanovPW12}. 

%--------------------------------------------------------
\subsubsection{Free-free absorption}

The quantization of electron motion gives rise to
cyclotron harmonics in the nonrelativistic
theory. \citet{PavlovPanov} derived photon
absorption cross-sections for an electron, which moves in a magnetic field and
interacts with a nonmoving point charge. This model is
applicable at $\omega \gg \omcp$. A more
accurate treatment of absorption of a photon by the system
of a proton and an electron yields (Paper~II;
\citealp{P10})
\begin{equation}
   \sigma_\alpha^\mathrm{ff}(\omega)
   =
          \frac{4\pi e^2
          }{ 
     \mel c} \,
   \frac{\omega^2\,\nu_{\alpha}^{\mathrm{ff}}(\omega)
          }{
          (\omega+\alpha\omc)^2 (\omega-\alpha\omcp)^2
             +\omega^2 \tilde\nu_\alpha^2(\omega)},
\label{sigma-fit0}
\end{equation}
where $\nu_{\alpha}^{\mathrm{ff}}$ is an effective
photoabsorption collision
frequency and $\tilde\nu_\alpha$ is a damping factor. In
the electron-proton plasma, taking into account the
scattering and free-free absorption, we have (Paper~II)
\begin{equation}
   \tilde\nu_\alpha =
     (1+\alpha{\omc}/{\omega})
     \nu_{\mathrm{i},\alpha}(\omega)
     + (1-\alpha{\omcp}/{\omega})
     \nu_{\mathrm{e},\alpha}(\omega)
       + \nu_{\alpha}^{\mathrm{ff}}(\omega).
\end{equation}
We see from
(\ref{sigma-fit0}) that
$\sigma_{-1}^\mathrm{ff}$ and
$\sigma_{+1}^\mathrm{ff}$ have a resonance at the
frequencies $\omc$ and $\omcp$, respectively. 
The effective free-free absorption frequency can be written
as 
\begin{equation}
      \nu_{\alpha}^{\mathrm{ff}}(\omega) =
        \frac{4}{3}\,\sqrt{\frac{2\pi}{\mel T}}\,
          \frac{n_\mathrm{e}\, e^4}{\hbar \omega}
           \Lambda_{\alpha}^{\mathrm{ff}}(\omega),
\label{nu-ff}
\end{equation}
where
$\Lambda_{\alpha}^{\mathrm{ff}}(\omega)$ is a dimensionless
Coulomb logarithm 
($\Lambda_{\alpha}^{\mathrm{ff}}=(\pi/\sqrt3)g_{\alpha}^{\mathrm{ff}}$,
where $g_{\alpha}^{\mathrm{ff}}$ is a Gaunt factor). Without
the magnetic field, $\Lambda_{\alpha}^{\mathrm{ff}}$ is a
smooth function of $\omega$. In a quantizing magnetic field,
however, it has peaks at the multiples of $\omc$ and $\omcp$
for all polarizations $\alpha$. An accurate calculation of
$\Lambda_{\alpha}^{\mathrm{ff}}(\omega)$ \citep{P10}
demonstrates that, unlike the electron cyclotron harmonics,
the ion cyclotron harmonics are so weak that they can be
safely neglected in the neutron-star atmosphere models.

In addition to the free-free absorption due to the
electron-proton collisions, in Paper~II we also considered
the free-free absorption due to the proton-proton
collisions. The results revealed that  at
$T\lesssim10^7$~K and $\hbar\omega/T\lesssim10$ the
corresponding photoabsorption cross section is much smaller
than the usual free-free absorption due to the
proton-electron collisions, whereas at larger $\hbar\omega$
it is smaller than the scattering cross section. Therefore
the proton-proton collisions can
be neglected in the opacity calculations.

%--------------------------------------------------------
\subsubsection{Bound-bound absorption}
\label{sect:bb}

%%%%%%%%%%%%%%%%%%%%%%%%%%%%%%%%%%%%%%%%%%%%%%%%%%%%%%%%%
\begin{figure*}
\includegraphics[height=.43\textwidth]{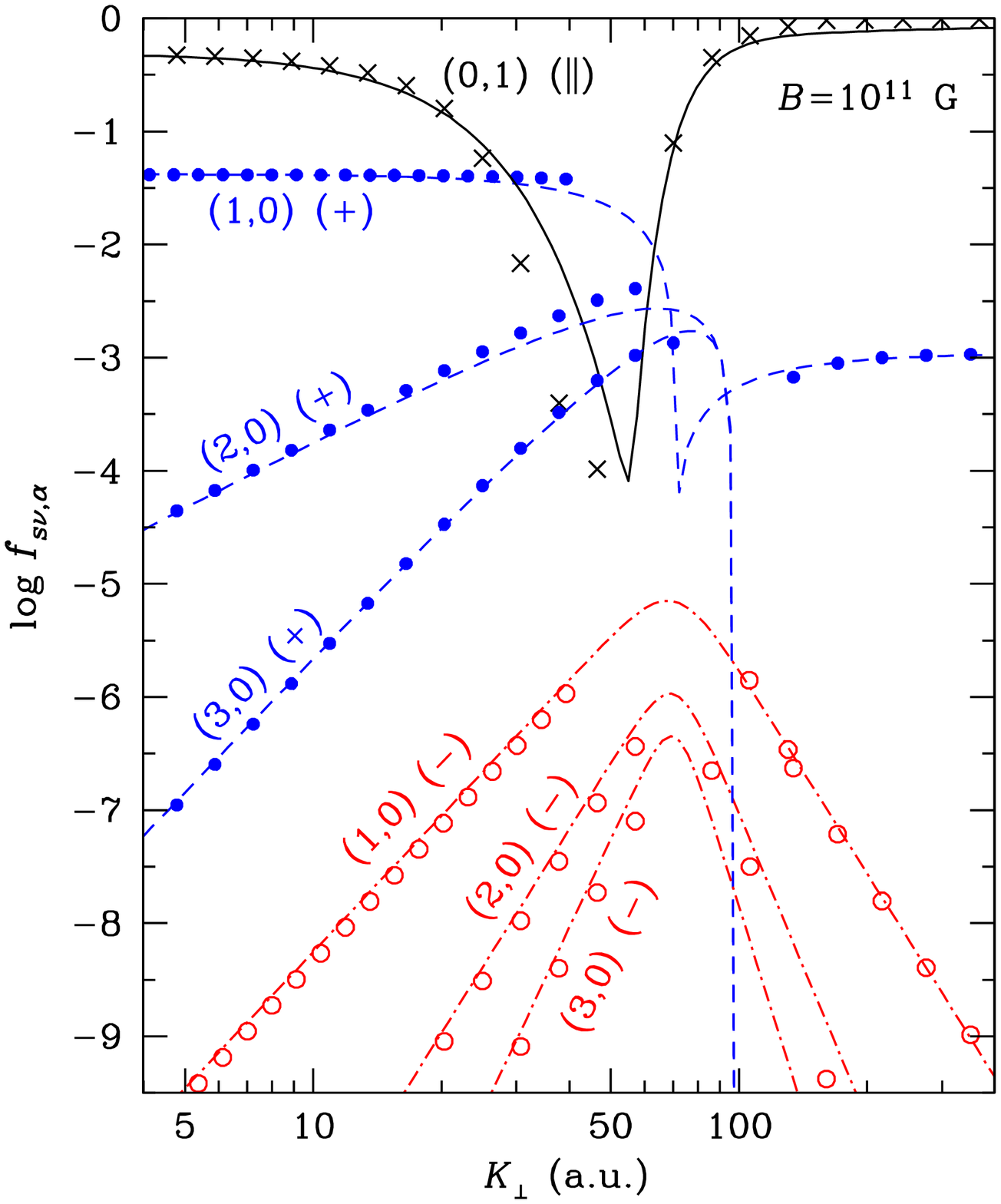}
\includegraphics[height=.43\textwidth]{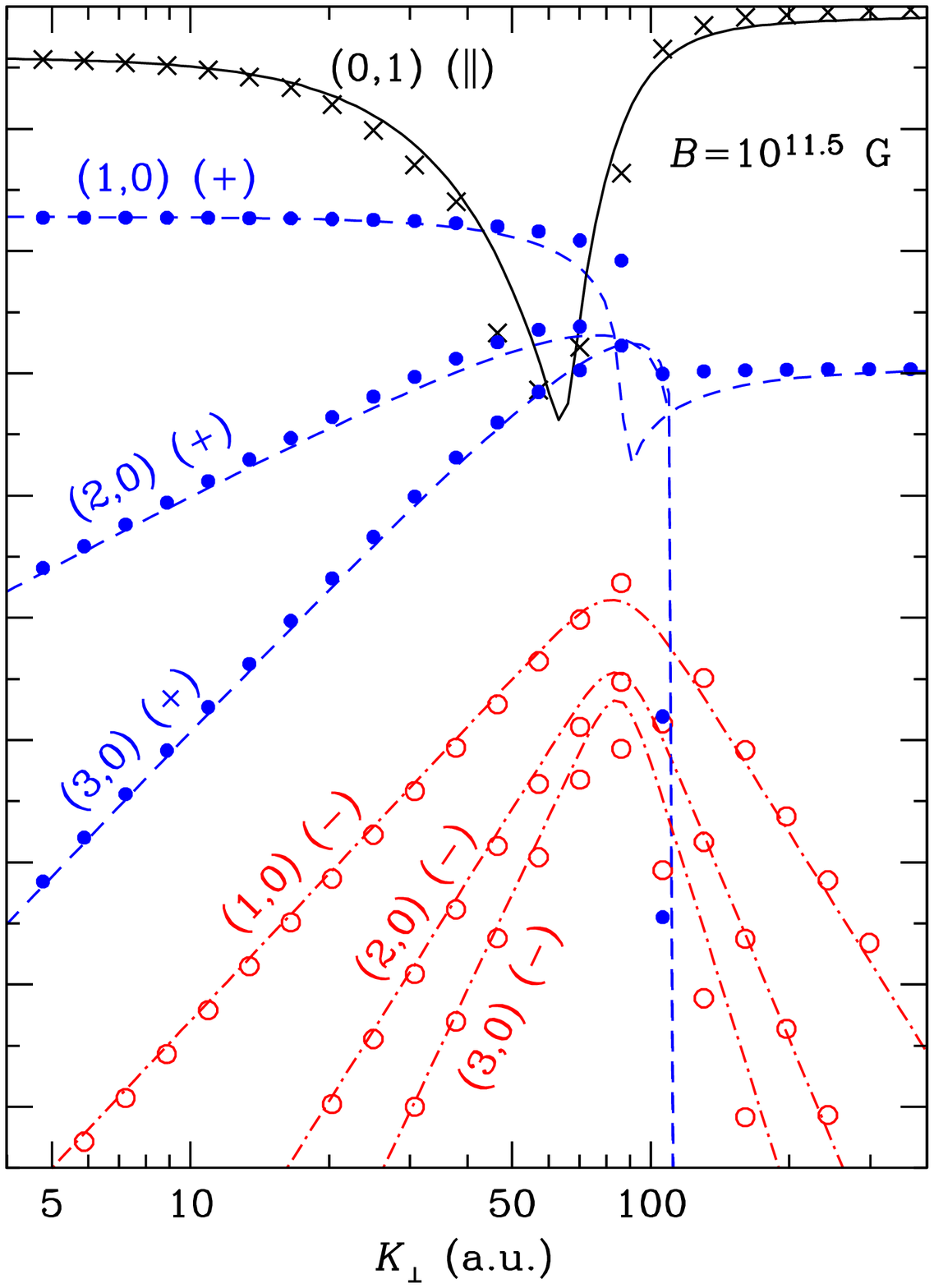}
\includegraphics[height=.43\textwidth]{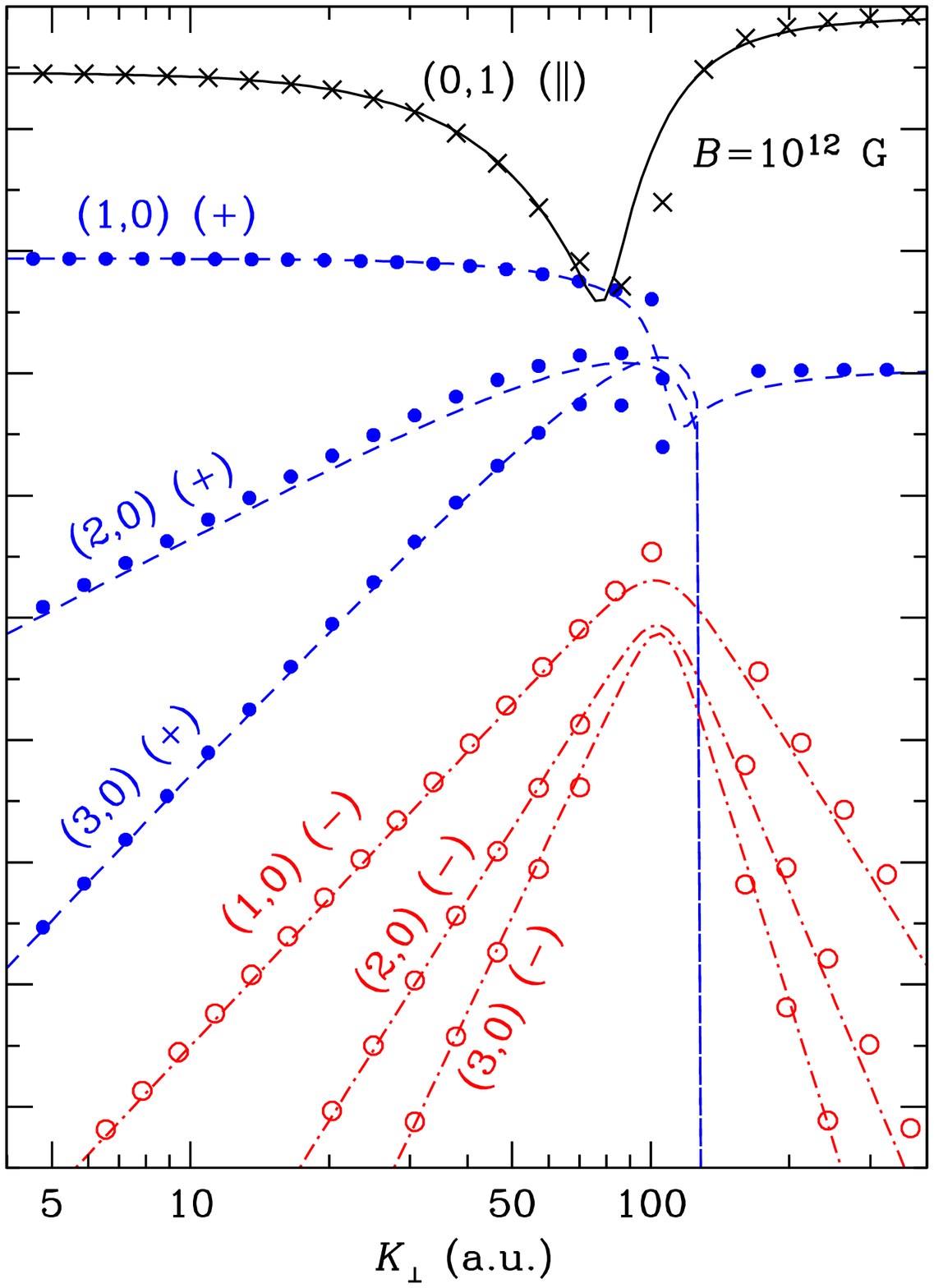}
\caption{Analytical approximations (lines) and results of
calculations (symbols) for oscillator strengths of
transitions from the ground state to different bound states
$(s,\nu)$ marked near the respective curves with absorption
of a photon with polarization $\alpha=-1$ (dot-dashed lines
and empty circles), $\alpha=0$ (solid lines and crosses), or
$\alpha=+1$ (dashed lines and filled dots) at $B=10^{11}$~G
(left panel), $3.16\times10^{11}$~G (middle panel), and
$10^{12}$~G (right panel). 
}
\label{fig:osc}
\end{figure*}
%%%%%%%%%%%%%%%%%%%%%%%%%%%%%%%%%%%%%%%%%%%%%%%%%%%%%%%%%

Bound-bound transitions of the H atom moving in a strong
magnetic field were studied  by \citet{PP95}. In the dipole
approximation, the cross section of an atom in a state
$(s,\nu)$ for absorption of a photon with frequency $\omega$
and polarization $\alpha$ with a transition to a state
$(s',\nu')$, averaged over the atomic states of motion,
reads
\bea
   \sigma_{\alpha;s\nu;s'\nu'}(\omega) &=& 
     \frac{2\pi^2e^2}{\mel c}
     \left(1-\mathrm{e}^{-\hbar\omega/T}\right)
   \frac{1}{A_{s\nu}}  \int_0^\infty 2\pi \Kp \dd \Kp 
\nonumber\\&&\times\,
         w_{s\nu}^\mathrm{o}(\Kp)
            \exp(E_{s\nu}(\Kp)/T)
\nonumber\\&&\times\,
   w_{s'\nu'}^\mathrm{o}(\Kp)
     f_{\alpha;s\nu;s'\nu'}(\Kp)
        \,\phi_{s\nu;s'\nu'}(\Kp,\Delta\omega),
\label{bb}
\eea
where 
\beq
   A_{s\nu} = \int_0^\infty {2\pi \Kp \dd \Kp 
         w_{s\nu}^\mathrm{o}(\Kp)
            \exp(E_{s\nu}(\Kp)/T)}
\eeq
is the normalization integral,
\beq
  f_{\alpha;s\nu;s'\nu'}(\Kp) = 
    \frac{\hbar\omega}{\mbox{Ry}}\,
  \left|
  \frac{\langle s',\nu',\Kp|r_{-\alpha}|s,\nu,\Kp\rangle}{\aB}
       \right|^2,
\label{f}
\eeq
is the oscillator strength in the dipole approximation,
$r_0=z$, $r_{\pm1}=(x\pm \mathrm{i}y)/\sqrt{2}$,
\beq
\Delta\omega=\omega-
 \left[E_{s\nu}(\Kp)-E_{s'\nu'}(\Kp)\right]/\hbar,
\eeq
and 
$\phi_{s\nu;s'\nu'}(\Kp,\Delta\omega)$ describes a
normalized profile of the spectral line at a fixed
$\Kp$, 
$
\int \phi_{s\nu;s'\nu'}(\Kp,\Delta\omega)
\dd\Delta\omega = 1.
$
The latter profile is assumed Lorentzian with  width
$\Gamma_{\alpha;s\nu;s'\nu'}(\Kp)$, determined by
electron-atom collisions \citep[see][]{PP95}. In
practice the collisional broadening plays minor role
compared to the magnetic broadening due to the dependence of
$\Delta\omega$ on $\Kp$, except for stationary
frequencies $\left[E_{s\nu}(\Kp)-E_{s'\nu'}(\Kp)\right]/\hbar$
for which $\dd E_{s\nu}(\Kp)/\dd \Kp =\dd
E_{s'\nu'}(\Kp)/\dd \Kp$ -- in particular, 
$\left[E_{s\nu}(\Kp)-E_{s'\nu'}(\Kp)\right]/\hbar$ at $\Kp\to0$,
$(s'-s)\omcp$ at $\Kp\to\infty$, and a small frequency
corresponding to anticrossings at $\Kp\sim\Kc$.
The magnetic broadening
exceeds by orders of magnitude the usual Doppler broadening,
which allows us to neglect the difference between the
pseudomomenta in the initial and final states in
Eqs.~(\ref{bb}) and (\ref{f}). The spectral
profile of the  bound-bound opacities becomes continuous in
a wide frequency range, resembling a reversed bound-free
profile.  We calculate the integral (\ref{bb}) using
analytical approximations for the  electron collision widths
$\Gamma_{\alpha;s\nu;s'\nu'}(\Kp)$
suggested in \citet{P98} and the approximations for binding
energies $E_{s\nu}(\Kp)$ and oscillator strengths
$f_{\alpha;s\nu;s'\nu'}(\Kp)$ presented in the
Appendix.

Examples of the oscillator strengths
are shown in Fig.~\ref{fig:osc}, where the
results of numerical calculations by the method described in
\citet{P94} are compared with the analytical approximations
(Appendix \ref{sect:osc}).
The figure shows oscillator strengths for the
main dipole-allowed transitions from the ground state to
excited discrete levels as functions of $\Kp$. Since the
atomic wave functions are symmetric with respect to the
$z$-inversion for the states with even $\nu$, and
antisymmetric for odd $\nu$, only the transitions that
change the parity of $\nu$ are allowed for the polarization
along the field ($\alpha=0$), and only those preserving the
parity for the circular polarizations ($\alpha=\pm1$). For
an atom at rest, in the dipole approximation, due to the
conservation of the $z$-projection of the total angular
momentum of the system, absorption of a photon with
polarization $\alpha$ results in the change of $s$ by
$\alpha$. This selection rule for a nonmoving atom
manifests itself in vanishing oscillator strengths at
$K_\perp\to0$ for $s\neq\alpha$. In an appropriate coordinate
system \citep{Burkova-ea76,P94}, the symmetry is restored at
$K_\perp\to\infty$, therefore the transition with $s=\alpha$ is
the only one that survives also in the limit of large
pseudomomenta. But in the intermediate region of $K_\perp$,
where the transverse atomic velocity is not small, the
cylindrical symmetry is broken, so that transitions to other
levels are allowed. For this reason the corresponding oscillator
strengths in Fig.~\ref{fig:osc} have maxima at
$K_\perp\sim \Kc$. 

%--------------------------------------------------------
\subsubsection{Bound-free absorption}

The theory of photoionization of the H atom in a strong
magnetic field with allowance for motion and a method of
calculation of the cross sections was described by
\citet{PP97}, who adapted the $R$-matrix formalism 
\citep{Rmatrix,Seaton83} to the case under study. Since this
numerical treatment is computationally involved, we first
compose a set of tables of the photoionization  cross
sections $\sigma_{\alpha;s\nu}^\mathrm{bf}(\omega,\Kp)$, and
then calculate the bound-free opacities using interpolation
across these tables. At a given $B$, for each of the three
basic polarizations $\alpha$, we calculate
$\sigma_{\alpha;s\nu}^\mathrm{bf}(\omega,\Kp)$ on a
predefined grid, as in Papers~II and~III. Now we have
refined the grid in photon energy, with
$\log_{10}\hbar\omega\ [\mathrm{eV}]$ ranging from 0 to 5
with step 0.01, and modified the grid of $\log_{10}\Kp$. In
order to avoid large numerical errors due to strong coupling
of the orbitals at $\Kp\sim\Kc$,
we exclude a range of $\Kp$ around $\Kc$ and use two
separate grids for $\Kp$ below and above the excluded
region. Each of these grids has, as previously, an equal
step in $\log \Kp$. The size of the excluded region is
determined ad hoc, from numerical tests at different $\Kp$ values.
The employed modification of the grid is justified by a
comparison of the results in the overlap range of
$B\sim10^{12}$~G, which reveals virtually no difference
between the opacities calculated with the old and
new tables of the bound-free cross sections.

As in Papers~II and~III, we filter out spurious outliers,
which appear because of the Beutler-Fano resonances, whose
widths are smaller than the step of our grid in
$\log\omega$, using the 3-point median filter at every
$\Kp$. The photoionization threshold is determined
independently for every $\Kp$, using the analytic fits to
the binding energies given in Appendix~\ref{sect:en}.

In addition to the bound-bound and bound-free atomic
transitions, in a plasma environment there are transitions
from bound states to the highly perturbed atomic states 
discussed in Sect.~\ref{sect:ioneq}. These perturbed levels
effectively dissolve and merge in a pseudo-continuum, which
lies below the photoionization threshold. In order to take
into account the radiative transitions into this
pseudo-continuum, we employ a technique of below-threshold
extrapolation,  which is usual in the zero-field case
\citep{DappenAM87,StehleJacquemot93,Seaton-ea94}. The
details of this technique for the case of an atom moving in
a strong magnetic field are given in Paper~II. 
As previously, the precalculated, filtered and extrapolated
photoionization cross sections are averaged over $\Kp$ with
statistical weights $w_{s\nu}^\mathrm{o}(\Kp)
\exp[E_{s\nu}(\Kp)/T]$. 

%--------------------------------------------------------
\subsection{Polarization vectors}
\label{sect:polar}

In the coordinate system $(x',y',z')$ rotated with respect to
$(x,y,z)$ so that $z'$ is along the wave vector $\bm{k}$ and
$\bm{B}$ is in the $(x',z')$ plane,
the electromagnetic polarization vectors $\bm{e}_j$
(Sect.~\ref{sect:modes}) can be written as \citep{HoLai01,HoLai03}
\beq
  (e_{j,x},e_{j,y},e_{j,z})=(1+K_j^2+K_{z,j}^2)^{-1/2} \,
  (i K_j, 1, i K_{z,j}),
\eeq
where 
\bea
&&
  K_j = \beta \left\{
   1 + (-1)^j \left[ 1 + \frac{1}{\beta^2} 
   + \frac{m}{1+{a}} \frac{\sin^2\theta_B}{\beta^2}\right]^{1/2}
   \right\},
\\&&
  K_{z,j} = - \frac{ 
   (\varepsilon_\perp' - \varepsilon_\|') K_j \cos\theta_B + \varepsilon_\wedge
   }{
   \varepsilon_\perp' \sin^2\theta_B + \varepsilon_\|' \cos^2\theta_B } 
   \, \sin\theta_B,
\\&&
 \beta = \frac{\varepsilon_\|' - \varepsilon_\perp' + \varepsilon_\wedge^2/\varepsilon_\perp' + \varepsilon_\|'
 \,{m}/(1+{a})
   }{
   2 \, \varepsilon_\wedge }
   \,\, \frac{ \varepsilon_\perp'}{\varepsilon_\|'}
   \,\,\frac{\sin^2\theta_B}{\cos\theta_B},
\\&&
\varepsilon_\perp' = \varepsilon_\perp + {a},
\qquad
\varepsilon_\|' = \varepsilon_\| + {a} +{q}.
\nonumber
\eea
Here, the parameters $K_j$ and $K_{z,j}$ are expressed in
terms  of the complex dielectric tensor of a plasma
(\ref{eps-p}),
dielectric tensor of vacuum  $4\pi \bm{\chi}^\textrm{vac} =
\textrm{diag}({a}, {a}, {a}+q )$, and the inverse magnetic
permeability  of the vacuum  $\bm{\mu}^{-1} = \mathbf{I} +
\textrm{diag}({a}, {a}, {a}+m)$. We calculate the plasma
dielectric tensor using the relation between the
polarizability coefficients $\chi_\alpha$ and the
opacities \citep{KK}
\begin{eqnarray}
  {\chi}^\mathrm{H}_\alpha(\omega) &=&
   \frac{c\rho}{4\pi^2\omega}\, \bigg\{\!
    \int_0^\omega \!\big[\,\kappa_\alpha (\omega+\omega')
     - \kappa_\alpha (\omega-\omega')\,\big]
     \frac{\dd\omega'}{\omega'}
\nonumber\\&
   + & \int_{2\omega}^\infty \frac{\kappa_\alpha(\omega')}{\omega'-\omega}
     \,\dd\omega'
   - \int_0^\infty \frac{\kappa_{-\alpha}(\omega')}{\omega'+\omega}
     \,\dd\omega' \bigg\} .
\label{KK-mu}
\end{eqnarray}
The vacuum polarizability and
permeability coefficients ${a}$, $q$, and $m$ can be
neglected at the relatively weak field strengths considered
here, but in general they have been fitted by elementary
functions \citep{KK}.

%%%%%%%%%%%%%%%%%%%%%%%%%%%%%%%%%
\section{Atomic signatures in the opacities and spectra}
\label{sect:results}

%%%%%%%%%%%%%%%%%%%%%%%%%%%%%%%%%%%%%%%%%%%%%%%%%%%%%%%%%
\begin{figure*}
\centering
\includegraphics[height=.43\textwidth]{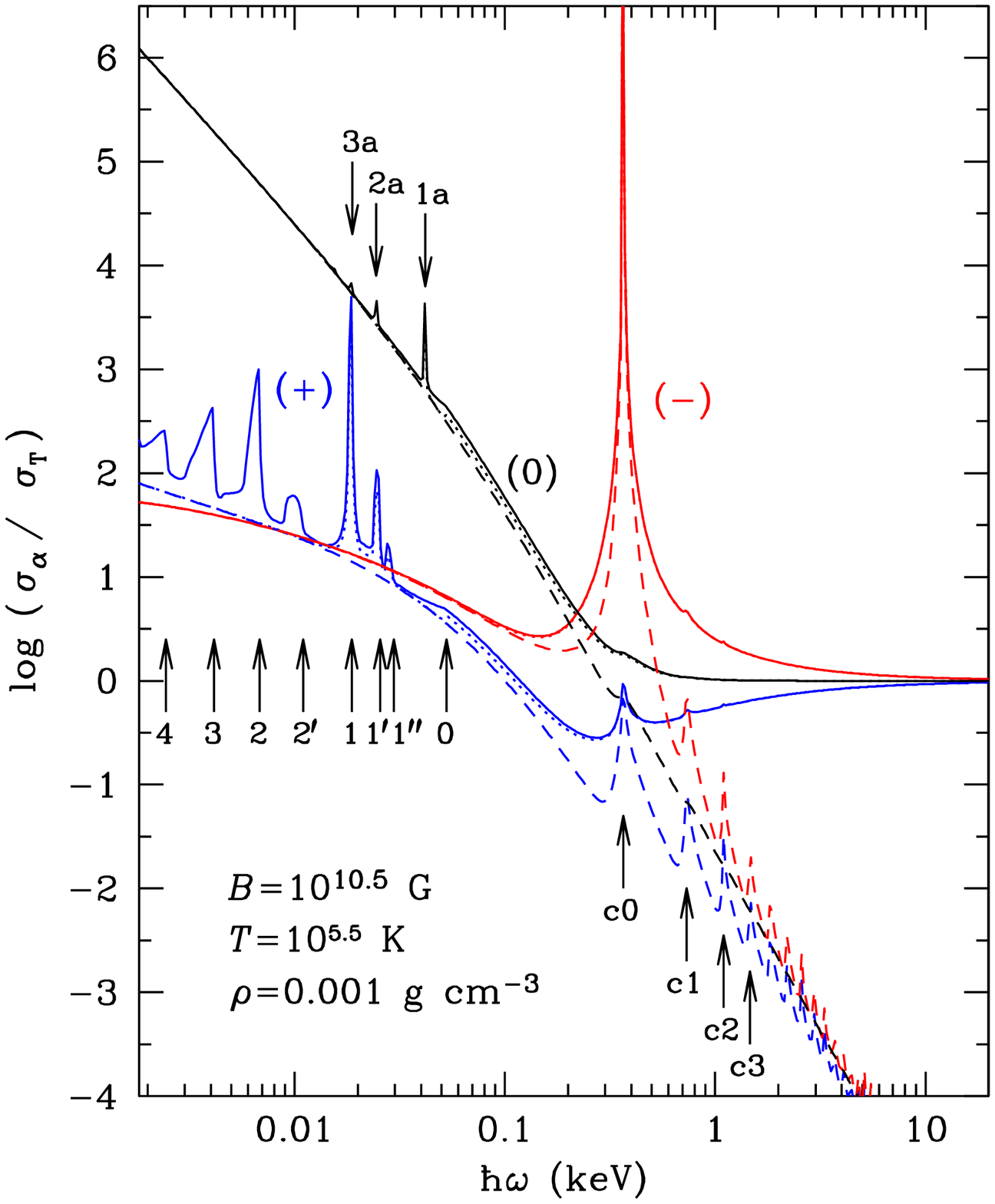}
\includegraphics[height=.43\textwidth]{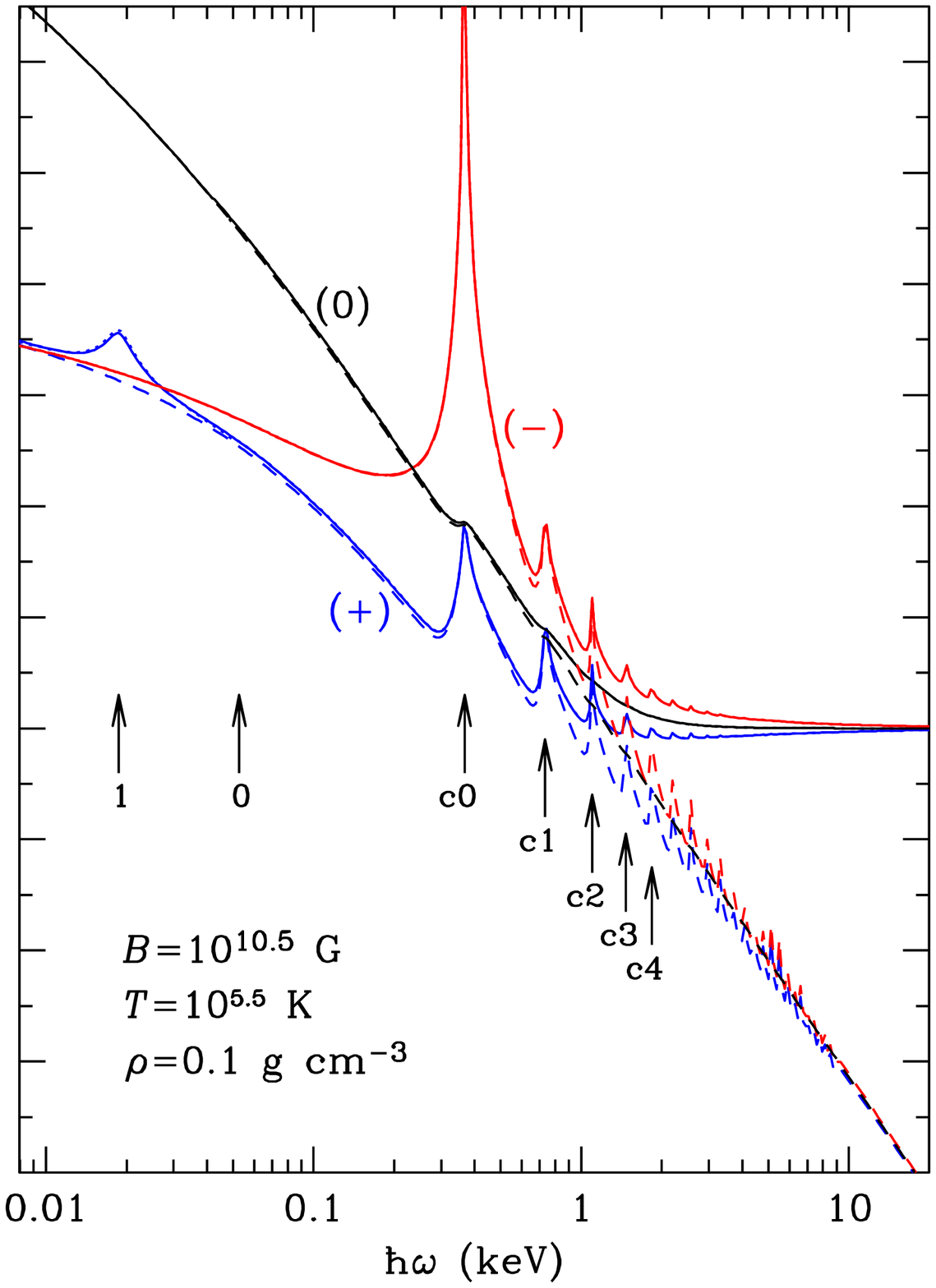}
\includegraphics[height=.43\textwidth]{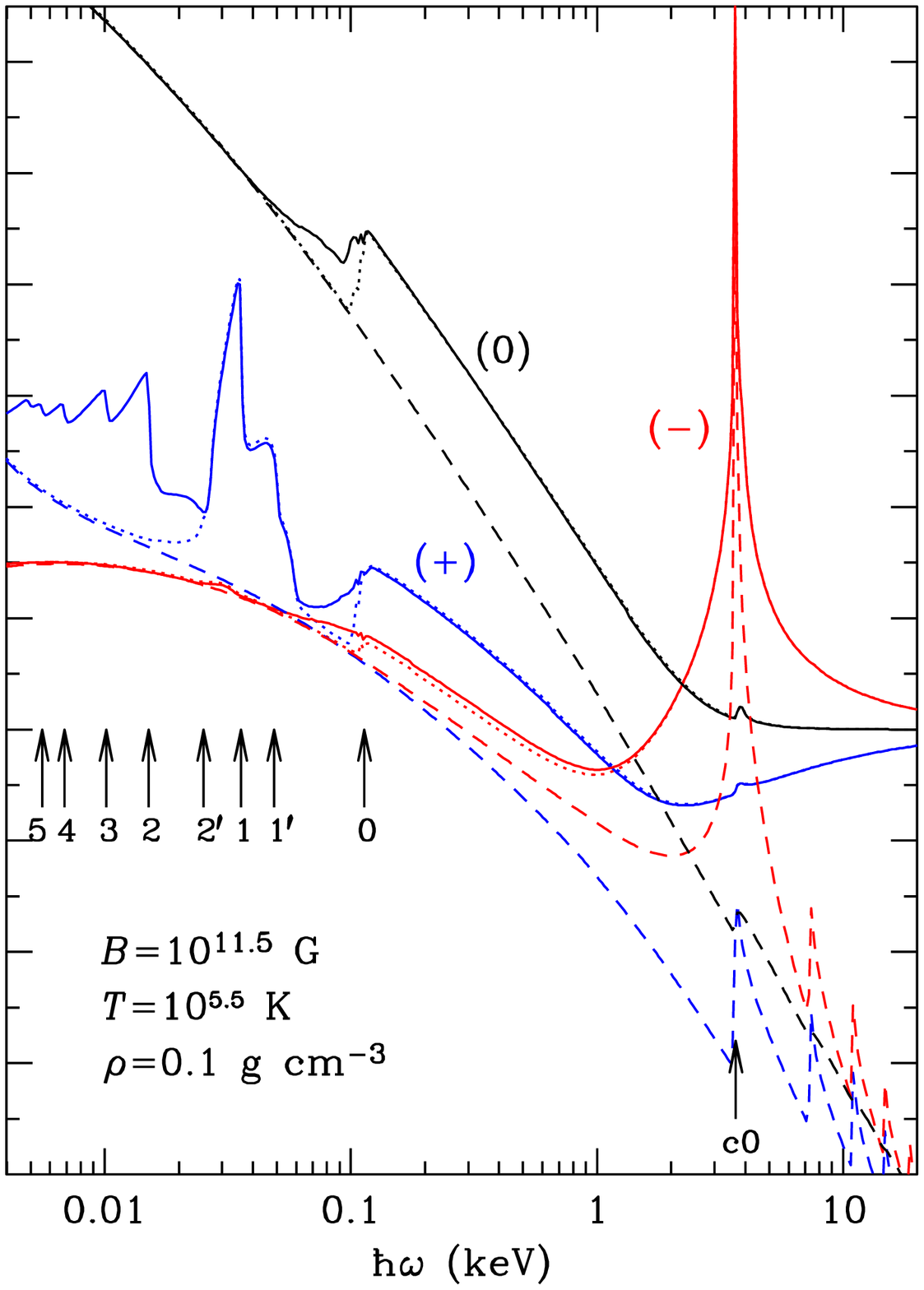}
\caption{Logarithms of total basic cross sections
$\sigma_\alpha$ (solid lines), $\alpha=-1,0,+1$ marked
``($-$)'', ``(0)'', and ''(+)'', respectively. For comparison,
the logarithms of cross sections without contribution of
excited bound states (dotted lines) and free-free cross
sections$\sigma_\alpha^\mathrm{ff}$ (dashed lines) are
plotted. The three panels correspond to three sets of plasma
parameters $\rho$, $T$, and $B$ marked on the plots. The
arrows correspond to different characteristic transition
energies for a nonmoving H atom (see text for discussion): 0
-- $E_{0,0}$ (the principal photoionization threshold), 
1 -- $(E_{0,0}-E_{1,0})$, $1'$ -- $(E_{0,0}-E_{2,0})$,
 $1''$ -- $(E_{0,0}-E_{3,0})$,
2 -- $(E_{1,0}-E_{2,0})$, $2'$ -- $(E_{1,0}-E_{3,0})$, 
3 -- $(E_{2,0}-E_{3,0})$, 
4 -- $(E_{3,0}-E_{4,0})$,
5 -- $(E_{4,0}-E_{5,0})$; 
1a -- $(E_{0,0}-E_{0,1})$, 2a -- $(E_{1,0}-E_{1,1})$, 3a --
$(E_{2,0}-E_{2,1})$. The arrows marked
c0, c1, c2, c3, c4 correspond to cyclotron harmonics
energies $(N+1)\hbar\omc$ with $N=0,1,2,3,4$, respectively.
}
\label{fig:sigma}
\end{figure*}
%%%%%%%%%%%%%%%%%%%%%%%%%%%%%%%%%%%%%%%%%%%%%%%%%%%%%%%%%

%--------------------------------------------------------
\subsection{Basic opacities}
\label{sect:opacbas}

Figure~\ref{fig:sigma} shows examples of the basic cross
sections $\sigma_\alpha$ in Eqs.~(\ref{kappa-a}),
(\ref{kappa-s}) at $T=3.16\times10^{5}$~K and different values of
$B$ and $\rho$. The left panel shows the case of
$B=3.16\times10^{10}$~G and relatively low density $\rho=10^{-3}$
\gcc. In this case, the number fraction of the atoms is
small, $x_\mathrm{H}=0.0025$. Nevertheless, we observe
prominent absorption features due to bound-bound transitions
between the neighboring tightly-bound states
$(s-1,0)\to(s,0)$ ($s=1,2,3,4$ marked in the figure), which
are allowed in the dipole approximation for $\alpha=+1$ for
a nonmoving atom. The lines marked $1'$ and $2'$ correspond
to the transitions $(s-1,0)\to(s+1,0)$ ($s=1,2$), which are
dipole-forbidden for an atom at rest, but become noticeable
for moving atoms. Since the energy difference
between the initial and final levels is smaller at
$\Kp\neq0$ than at $\Kp=0$ (cf.{}
Fig.~\ref{fig:enfit14}), the maxima of the spectral lines are
shifted to the left from the corresponding arrows in the
figure, which are plotted at
$\hbar\omega=E_{s+1,0}(0)-E_{s-1,0}(0)$.  For the
longitudinal polarization $\alpha=0$, there are narrow
spikes, marked 1a, 2a, 3a, to the left of the energy
$E_{00}(0)$, which are due to the transitions from the
tightly-bound states to the lowest loosely-bound states of
the same $s$-manifold, $(s,0)\to(s,1)$, with $s=0,1,2$,
respectively. At higher energies, we see the peaks marked c0, c1, c2, c3
at the dashed curves. They are due to the absorption at the
cyclotron fundamental frequency $\omc$ and its harmonics.
The harmonics are not seen, however, in the total cross
sections (solid lines in the left panel), because they are
dominated by scattering at these energies and plasma
parameters. 

The quantum cyclotron harmonics become visible in the total
cross sections at higher density $\rho=0.1$ \gcc{} (the
middle panel), because of the larger free-free absorption.
Although the abundance of the atoms is also larger,
$x_\mathrm{H}=0.021$, the atomic absorption features are
merged into the free-free continuum. The absorption features
due to transitions between neighboring tightly-bound states
reappear at a stronger field $B=3.16\times10^{11}$~G (the
right panel), partly because of a higher abundance of the
atoms ($x_\mathrm{H}=0.053$), but mainly because of the
lowering of the continuum level for $\alpha=-1$ with
increasing $B$. In the latter case, the cyclotron-absorption
harmonics are again submerged under the scattering. The
signatures of bound-bound transitions with absorption of a
photon polarized along $\bm{B}$ ($\alpha=0$) are not visible
in the middle and right panels, because the loosely-bound
states are destroyed by pressure ionization and form
quasicontinuum at this density.

The ground-state photoionization jump above the free-free
continuum at $\hbar\omega=E_{00}$ is small at $B=3.16\times10^{10}$~G
in the left panel of Fig.~\ref{fig:sigma} and virtually
absent in the middle panel, because the product
$x_\mathrm{H}\sigma_\alpha^\mathrm{bf}$ is smaller than
$\sigma_\alpha^\mathrm{ff}$ in these cases. However, it is
clearly visible at the higher field strength
(the right panel). This jump is smoothed by the magnetic
broadening and by photoionization
of excited tightly-bound states, as seen from a
comparison with the model where the excited states are
neglected, which is plotted by the dotted lines.

%%%%%%%%%%%%%%%%%%%%%%%%%%%%%%%%%%%%%%%%%%%%%%%%%%%%%%%%%
\begin{figure*}
\centering
\includegraphics[height=.43\textwidth]{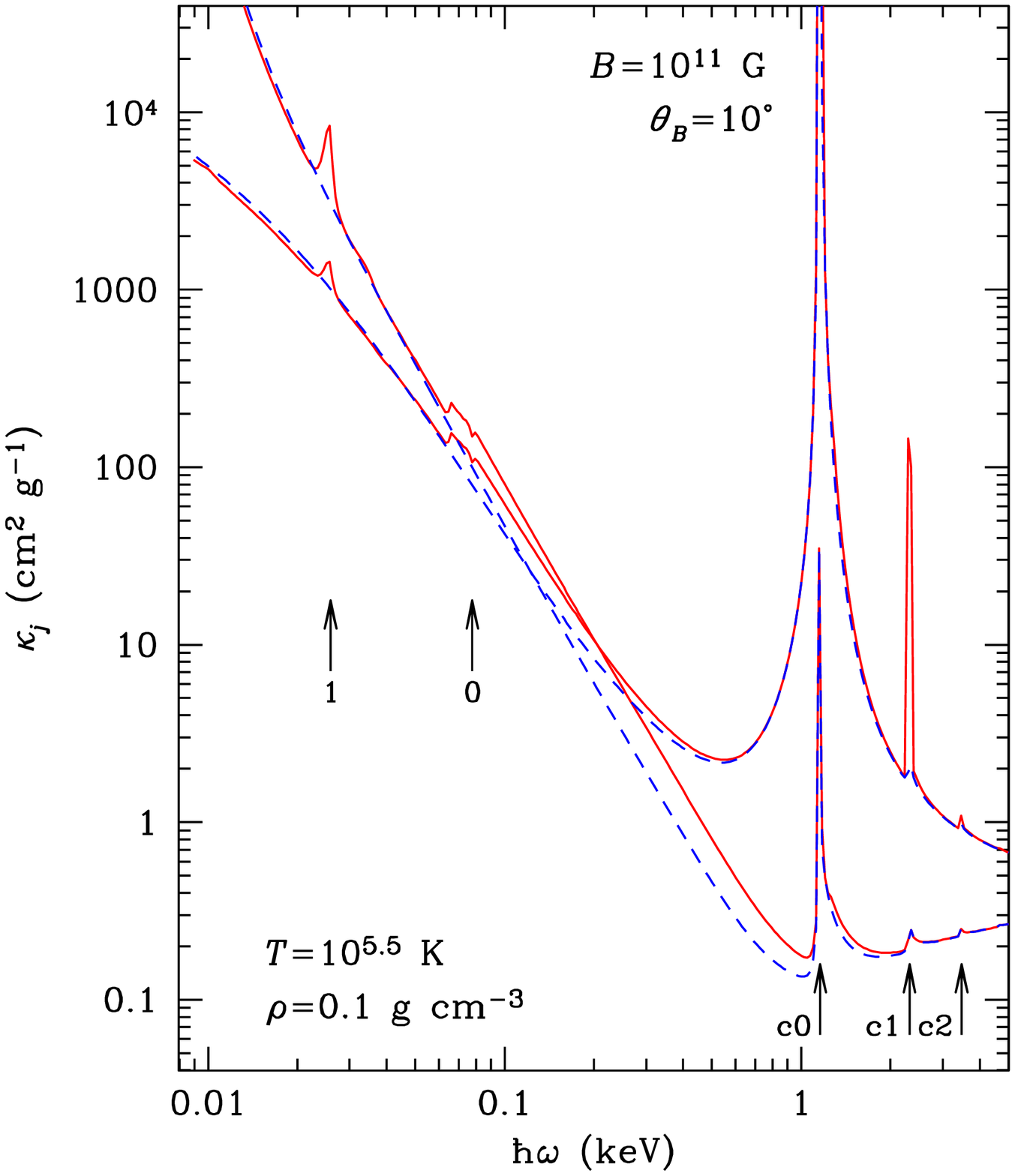}
\includegraphics[height=.43\textwidth]{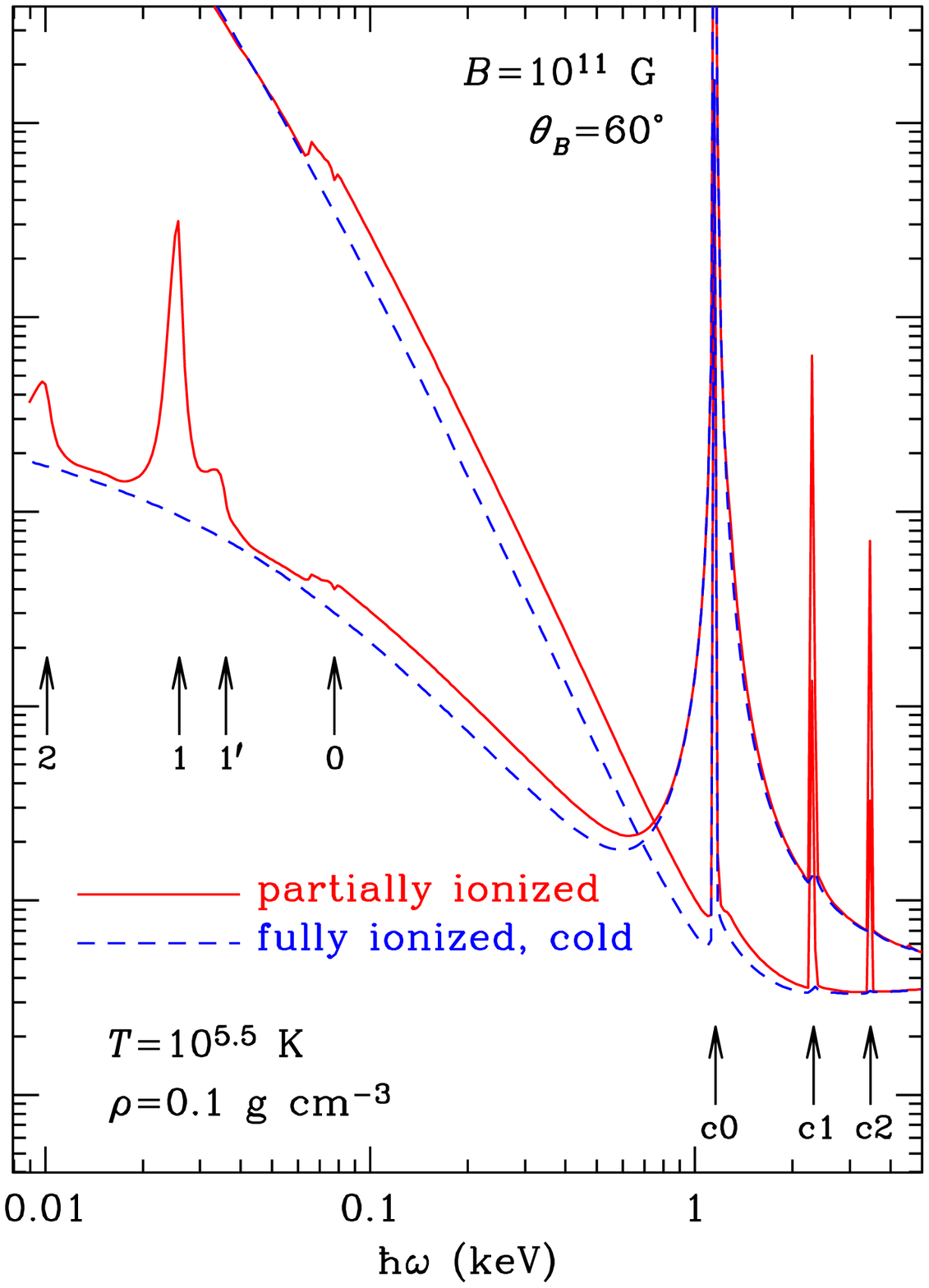}
\includegraphics[height=.43\textwidth]{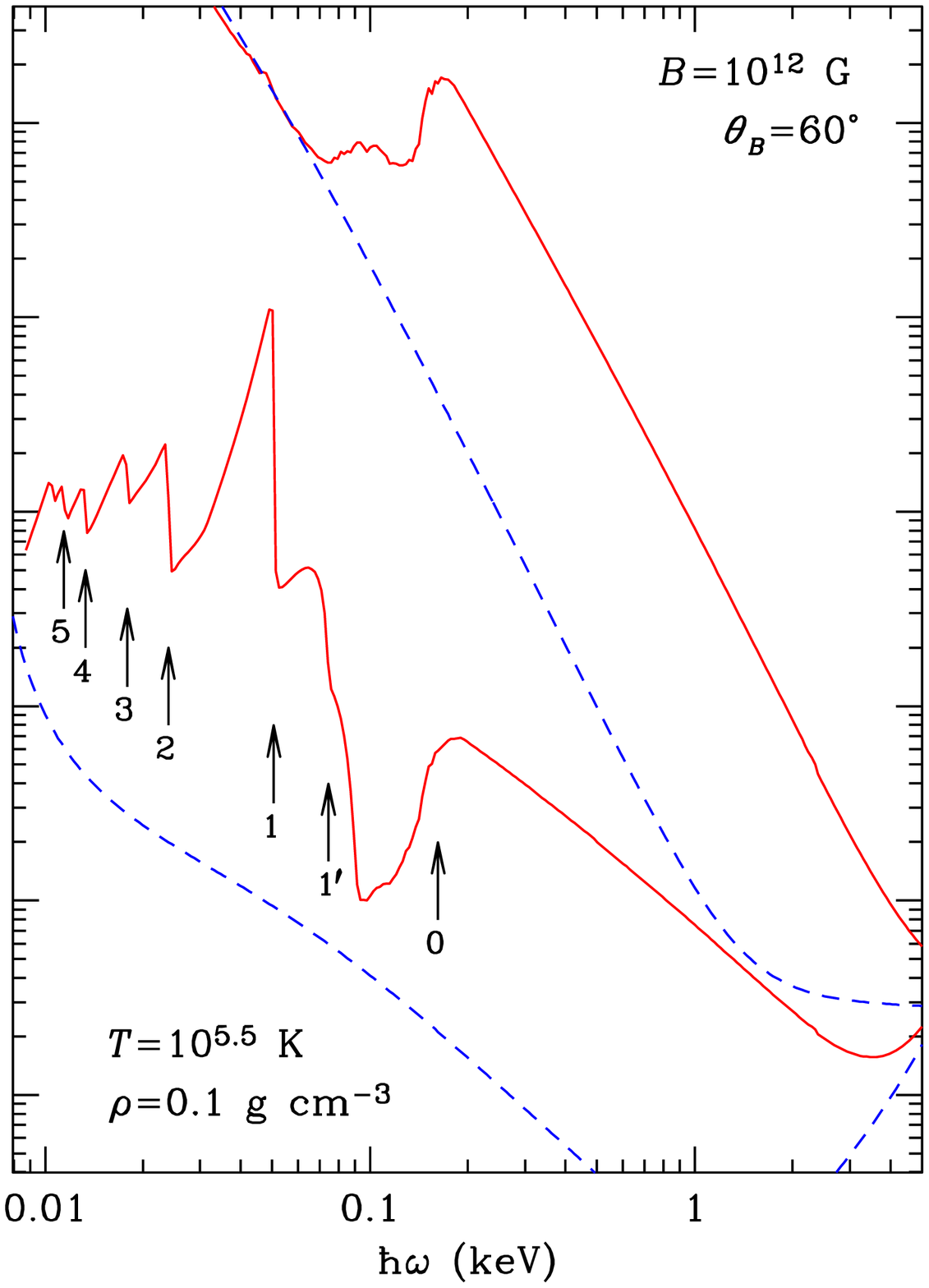}
\caption{Opacities $\kappa_j$ of the normal polarization
modes $j=1,2$ (the lower and upper curve of each type,
respectively) in a magnetized plasma at $\rho=0.1$ \gcc{}
and $T=10^{5.5}$~K, for $B=10^{11}$~G and
$\theta_B=10^\circ$ (left panel), $B=10^{11}$~G and
$\theta_B=60^\circ$ (middle panel), and $B=10^{12}$~G and
$\theta_B=60^\circ$ (right panel). Accurate opacities (solid
lines) are compared with the approximation of the cold,
fully ionized plasma (dashed lines).
The arrows correspond to different characteristic transition
energies and are marked in the same way as in
Fig.~\ref{fig:sigma}.
}
\label{fig:opac}
\end{figure*}
%%%%%%%%%%%%%%%%%%%%%%%%%%%%%%%%%%%%%%%%%%%%%%%%%%%%%%%%%

%--------------------------------------------------------
\subsection{Opacities for the normal modes}
\label{sect:opacnorm}

The basic opacities obtained in Sect.~\ref{sect:opacbas}
have been used to calculate plasma polarizabilities and
polarization vectors of the normal modes
(Sect.~\ref{sect:polar}).

Figure~\ref{fig:opac} illustrates the effect of incomplete
ionization on the opacities for the two normal modes
(Sect.~\ref{sect:modes}). Here, the atomic features
analogous to those in Fig.~\ref{fig:sigma} are also seen. 
The features marked by numbers 2 through 5 arise from the
bound-bound radiative transitions from excited tightly bound
states, which were not included in the previous opacity
calculations \citep{KK}. In addition to the bound-bound,
bound-free, and free-free absorption,  we have also included
the cyclotron absorption beyond the cold plasma
approximation, following \citet{SuleimanovPW12}. The latter
absorption gives rise to the high and narrow peaks, marked
``c1'' and ``c2'' in the left and middle panels. These peaks
correspond to the synchrotron harmonics \citep{Zheleznyakov}
and thus they present a manifestation of an effect of
special relativity. For comparison, we plot by dashed lines
the opacities calculated in the approximation of cold, fully
ionized plasma. In the latter approximation, the atomic
features are absent because of the full ionization,  and the
peaks  at the cyclotron harmonics are much smaller. The
latter difference demonstrates that, despite the smallness
of the relativistic parameters $T/\mel c^2$ and
$\hbar\omc/\mel c^2$, the relativistic effects substantially
change the opacities at the cyclotron harmonics frequencies,
in agreement \citet{SuleimanovPW12}.

Figures~\ref{fig:sigma} and \ref{fig:opac} show that
photoionization becomes substantial at relatively strong
magnetic fields. The contribution of the bound-bound
transitions into the opacities also increases with field
increase, but the bound-free absorption grows faster and
becomes more important. This tendency continues at higher
fields, so that the bound-bound transitions becomes
unimportant for magnetars, at contrast to the bound-free
ones (Paper~III).

%--------------------------------------------------------
\subsection{Spectra}

%%%%%%%%%%%%%%%%%%%%%%%%%%%%%%%%%%%%%%%%%%%%%%%%%%%%%%%%%
\begin{figure}
\centering
\includegraphics[width=\columnwidth]{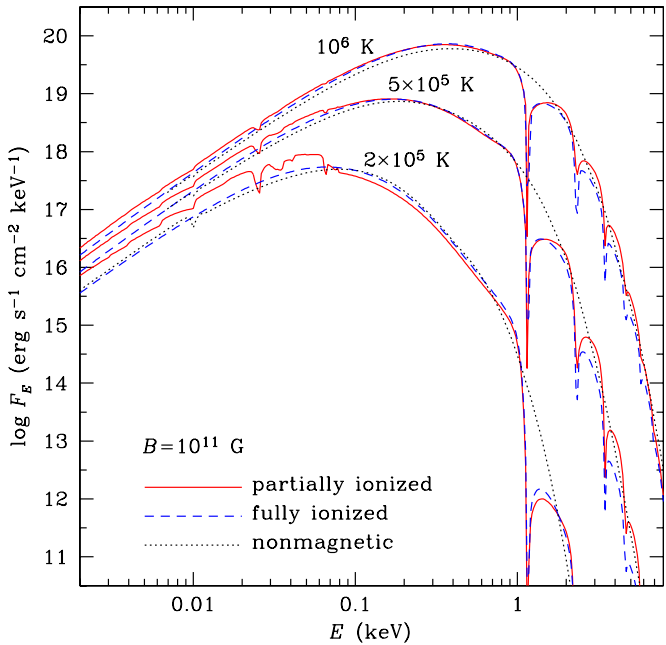}
\caption{The local spectrum of a neutron star with magnetic
field $B=10^{11}$~G, normal to the surface, and with effective
temperatures $T_\mathrm{eff}=10^6$~K, $5\times10^5$~K, and
$2\times10^5$~K (marked near the curves). The partially ionized
atmosphere spectrum (solid lines) is compared with the fully ionized
atmosphere model (dashed lines) and with the nonmagnetic
atmosphere model (dotted lines).
}
\label{fig:sp11}
\end{figure}
%%%%%%%%%%%%%%%%%%%%%%%%%%%%%%%%%%%%%%%%%%%%%%%%%%%%%%%%%

%%%%%%%%%%%%%%%%%%%%%%%%%%%%%%%%%%%%%%%%%%%%%%%%%%%%%%%%%
\begin{figure}
\centering
\includegraphics[width=\columnwidth]{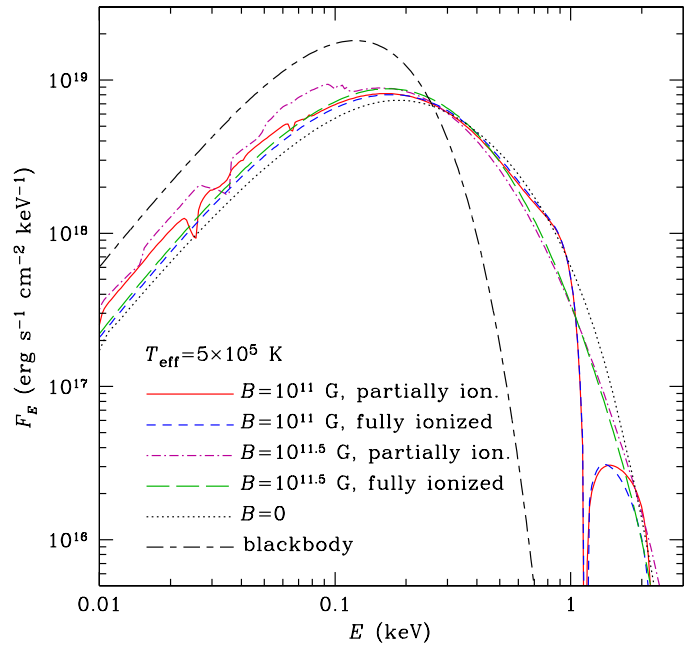}
\caption{The local spectrum of a partially ionized
atmosphere of a neutron star with
effective
temperature $T_\mathrm{eff}=5\times10^5$~K, with magnetic
fields $B=3.16\times10^{11}$~G
(dot-dashed line), $10^{11}$~G (solid line) and 0~G (dotted
line). Blackbody spectrum (short-long-dashed line) and
spectra of a magnetized, but fully ionized, atmosphere are also shown.
}
\label{fig:spt5.7}
\end{figure}
%%%%%%%%%%%%%%%%%%%%%%%%%%%%%%%%%%%%%%%%%%%%%%%%%%%%%%%%%

We have included the calculated opacities in the  equations
of radiative transfer for the two normal modes and solved it
numerically with the equations of hydrostatic and 
energy balance, using the numerical method developed by
\citet{HoLai01}. Examples of the resulting atmosphere
spectra are shown in Figs.~\ref{fig:sp11} and
\ref{fig:spt5.7}.

Figure \ref{fig:sp11} shows a spectrum of a neutron star
with magnetic field $B=10^{11}$~G and effective temperatures
$T_\mathrm{eff}=10^6$~K, $5\times10^5$~K, and
$2\times10^5$~K,  calculated using the models of fully and
partially ionized hydrogen atmospheres. In these examples,
we assumed gravity $g=2\times10^{14}$ cm s$^{-2}$, which
corresponds, for example, to a neutron star with mass
$M=1.4\,M_\odot$ and radius $R=10.9$ km. For realistic
neutron-star equation of state BSk21
\citep{Goriely-ea10,Potekhin-ea13}, this $g$ corresponds to
$M=1.8\,M_\odot$ and $R=12.5$~km. The first and second
values of the effective temperatures fall in the range of
current observational estimates for a number of thermally
emitting neutron stars \citep{Vigano-ea13}, albeit the hot
spots observed on CCOs have $T>10^6$~K.  The third value,
$T_\mathrm{eff}=2\times10^5$~K,  has not been observed.
Indeed, thermal radiation of such cold neutron star is
difficult to measure because of the low thermal flux.
However, this value of $T_\mathrm{eff}$ is also plausible,
and if there are such neutron stars at distances within
$\sim100$~pc, their thermal spectra may be measured in the
future. For the partially ionized models with
$T_\mathrm{eff}\geq5\times10^5$~K, one can notice the
absorption feature at $E=26$~eV, which corresponds to the
transition between the ground state and the lowest excited
state of the H atom, but otherwise the spectrum is smooth
and close to the one in the fully ionized plasma model. The
atomic features are rather small and less significant than
the cyclotron harmonics.  At $T_\mathrm{eff}=2\times10^5$~K,
more atomic spectral features are discernible. At this field
strength, they lie at rather low energies $E\lesssim0.1$
keV.

Figure \ref{fig:spt5.7} presents a comparison of the spectra
for $B=10^{11}$~G and
$3.2\times10^{11}$~G at $T_\mathrm{eff}=5\times10^5$~K,
calculated using the models of fully and partially ionized
hydrogen atmosphere. At
the stronger field, the atomic absorptions make a noticeable
contribution to the spectrum at energies
$\hbar\omega\sim10$\,--\,100 eV, but even in this case they
are not well pronounced. We conclude that the atomic
absorption is not very important for the atmosphere spectra
at $B\lesssim10^{11}$~G, provided that
$T_\mathrm{eff}\gtrsim5\times10^5$~K.

In Figs.~\ref{fig:sp11} and \ref{fig:spt5.7} we also show
the spectra in the model of partially ionized nonmagnetic H
atmosphere (for details of this calculation, see
\citealp{HoHeinke09}). In the latter model, there is no
cyclotron lines, atomic
absorption features are very weak (barely discernible)
because of a high degree of plasma ionization, and
the spectral maximum is shifted to higher energies by
$\sim10\%$ compared with the case of $B=10^{11}$~G. 
For comparison, in Fig.~\ref{fig:spt5.7} we plot
the blackbody model, which strongly underestimates the peak
energy and overestimates the peak flux.

%%%%%%%%%%%%%%%%%%%%%%%%%%%%%%%%%
\section{Conclusions}
\label{sect:concl}

We have developed new analytical approximations for
energies, sizes, and oscillator strengths of a H atom moving
arbitrarily in moderately strong magnetic field
$3\times10^{10}\mbox{~G}\lesssim B\lesssim10^{12}$~G. Using
these approximations and extensive numerical calculations of
the bound-free absorption cross sections, we calculated the
ionization equilibrium, equation of state, and opacities at
the moderate fields. The tables of the thermodynamic
functions, atomic fractions,  and Rosseland opacities,
previously available online for $B=10^{12}$\,--\,$10^{15}$~G
(Papers~I and II), are supplemented by the field interval
$B=3\times10^{10}$\,--\,$10^{12}$~G.

The calculated spectral opacities are implemented in
calculations of the neutron-star atmosphere models. The
results show that at  $B\ll10^{12}$~G and
$T_\mathrm{eff}\gtrsim5\times10^5$~K the atomic absorption
features in the spectra are small. Bound-bound features at
such field strengths are more significant than bound-free
ones, but they occur at low energies, which are difficult to
observe. Moreover, the distribution of the magnetic field
over the surface should additionally smear these features
and make them less significant, as has been demonstrated in
the case of the stronger fields by \citet{HoPC08}. On the
other hand, despite the smallness of the characteristic
thermal and photon energies compared to the electron rest
energy, the relativistic cyclotron harmonics are clearly
visible in the spectra at these field strengths, in
agreement with the results previously reported by
\citet{SuleimanovPW12} and \citet{Ho13}.

\begin{acknowledgements}
The work of AYP on calculation of polarization modes and
opacities (Sect.~4) has been supported  by the Russian Science
Foundation (grant 14-12-00316). 
WCGH appreciates use of computer facilities
at KIPAC.
\end{acknowledgements}

%%%%%%%%%%%%%%%%%%%%%%%%%%%%%%%%%%%%%%%%%%%%%%%%%%
\appendix
\section{Analytical approximations for atomic energies,
sizes, and oscillator strengths}

% ------------------------------------------
\subsection{Binding energies}
\label{sect:en}

For the tightly bound states  of a nonmoving H atom (any
$s$, $\nu=0$, $\Kp=0$), we use the analytical approximations
for binding energies from \citet{P14}.  For the loosely
bound states  of a nonmoving H atom ($\nu\geq0$, $\Kp=0$),
we use the analytical approximations from \citet{P98}. Both
sets of approximations are valid for many quantum states at
$\gamma\gtrsim1$.

For moving atoms ($\Kp>0$) we use different approximations
for the centered states, $E_{s\nu}^{(<)}(\Kp)$, and
decentered states, $E_{s\nu}^{(>)}(\Kp)$, and replace the
inflection point at $\Kc$ by the intersection of these two
functions. For the centered states, we use \req{E-small-K}
with
\beq
   m_{s\nu}=\mH\,[1+(\gamma/\gamma_{s\nu})^{p_{s\nu}}]
\eeq
where 
$\gamma_{s\nu}$ and $p_{s\nu}$ are dimensionless parameters,
which are approximated as functions of $s$ and $\nu$:
\bea&&\hspace*{-1em}
   \gamma_{s0} = 6\times10^3/(1+2s)^2,
\quad
    p_{s0}=0.9,
\nonumber\\&&\hspace*{-1em}
   \gamma_{s\nu} = \frac{110}{n^2\,(2-s+s^2)},
\quad n=\frac{\nu+1}{2},
\quad p_{s\nu}=1.65
 \quad (\mbox{odd~}\nu),
\nonumber\\&&\hspace*{-1em}
   \gamma_{s\nu} = \frac{55}{n^2\,(1+s+s^2)},
\quad n=\frac{\nu}{2},
\quad p_{s\nu}=1.2
 \quad (\mbox{even~}\nu\geq2).
\nonumber\eea
For the decentered states with $\nu=0$, we use
\req{E-large-K} with
\beq
   \epsilon_{s0} = \hat{r}_*/(5+3s) +
      (2\mbox{Ry}/E_{s0}^\|)^2.
\eeq
For the decentered states with $\nu\geq1$,  we find that the
formulae used previously for the binding energies at any
$\Kp$ for $\gamma>300$ \citep{P98}
remain valid at smaller $\gamma$ for $\Kp>\Kc$.

These approximations are sufficiently accurate for modeling
neutron star photospheres. The differences between
neighboring energy levels are determined by these formulae
with an accuracy of a few percent (except the $\Kp$ ranges
near anticrossings) at $10\lesssim\gamma\lesssim1000$.
Examples of calculated and fitted binding energies are shown
in Fig.~\ref{fig:enfit14}.

% ------------------------------------------
\subsection{Atomic sizes}

For the centered states, the electron cloud is cylindrically
symmetric at $\gamma\gg1$, except for the ranges of $\Kp$
near level anticrossings. The root-mean-square (rms)
sizes of this cloud transverse to the magnetic field are
\beq
   l_{x}=l_{y}=\aB\,\sqrt{(s+1)/\gamma},
\quad
   l_{\perp}=
         \aB\,\sqrt{2(s+1)/\gamma}.
\eeq
The atomic size along the field
is given at $\Kp=0$ by the approximation
\citep{P98}
\bea&&
   l_{z}^{(0)} = \frac{\aB}{\sqrt{2}} +
\frac{\aB}{\ln[\gamma/(1+s)]}\,\frac{\mbox{Ry}}{E_{s0}^{\|}(0)}
\quad(\nu=0),
\\&&
  l_{z}^{(0)} = 1.6 \,\aB (\mbox{~Ry}/E_{s\nu}^{\|}(0))
\quad(\nu\geq1).
\label{lzcentr}
\eea
This size remains almost constant for the centered states of
a moving atom. 
For the decentered states ($\Kp>\Kc$), our approximation of
the longitudinal size reads
\beq
  l_{z} = \left[
   \aB^2 \left(\nu+1/2\right)
      \sqrt{\hat{r}_*^3
      +(4.3+7\nu^2)\,\hat{r}_*^2}
        + \left(l_{z}^{(0)}\right)^2 \right]^{1/2}
\label{lzdecentr}
\eeq
Unlike the case of $\gamma>300$ \citep{P98}, 
at smaller $\gamma$ we do not smooth the transition between
\req{lzcentr} and \req{lzdecentr}.

Although the electron cloud is mostly cylindrically
symmetric, the atom is not, because the proton and electron
are not centered at the same axis. The atom acquires a
constant dipole moment proportional to the mean
electron-proton separation $\bar{x}$, which is considerably
smaller than $r_*$ for the tightly bound states at
$\Kp<\Kc$  and approaches $r_*$ for the loosely bound or
decentered states. At $\gamma\gtrsim10$, the fractional
difference between $\bar{x}$ and $r_*$ can be approximately
described by expressions \beq 1- \frac{\bar{x}}{r_*} = 
\left\{ \begin{array}{l}\displaystyle
\frac{1}{1+0.0014(1+s)^2 \,\gamma^{0.77}} \quad(\nu=0, ~
\Kp<\Kc), \\[3ex]\displaystyle \frac{1}{1+(\gamma/55)^{3/2}}
\quad (\nu=1,2, ~ \Kp<\Kc), \\[3ex]\displaystyle
\left(\frac{17\ln(1+\gamma)\,(1-(1.5+2s)^{-1})}{ \gamma
\hat{r}_*} \right)^3 ~~~ (\nu=0, ~\Kp>\Kc), \\[3ex] 0 \quad
\mbox{otherwise.} \end{array} \right.\hspace*{-3em} \eeq
Then the total rms size that is used for calculation of the
occupation probabilities (Paper~I) is given by $
\left(\bar{x}^2+l_x^2+l_y^2+l_z^2\right)^{1/2}. $

% ------------------------------------------
\subsection{Oscillator strengths}
\label{sect:osc}

In this section we present analytical approximations to the
oscillator strengths $f_{\alpha;s\nu;s'\nu'}(\Kp)$,
discussed in Sect.~\ref{sect:bb}. 
It is sufficient to retain only 
the transitions with $\nu'=\nu\pm1$ for $\alpha=0$ and with
$\nu'=\nu$ for $\alpha=\pm1$, because the other oscillator
strengths are very small due to the smallness of the
wave-function overlap
integral implied in the matrix element in \req{f}. As can be
seen in Fig.~\ref{fig:ie}, the
loosely-bound states are populated very weakly
compared to the tightly bound states, therefore we can
restrict the consideration by initial states
with $\nu=0$. Furthermore, the symmetry relation
$f_{\alpha;s\nu;s'\nu'}=f_{-\alpha;s'\nu';s\nu}$ allows us
to consider only the cases where $s'\geq s$. 
Thus we are left with oscillator strengths
$f^{(\alpha)}_{s,\Delta s}=f_{\alpha;s,0;s+\Delta s,0}$ 
for $\alpha=\pm1$
and
$f^{(0)}_s=f_{0;s,0;s,1}$ for $\alpha=0$.

In the dipole
approximation for the nonmoving atom, the only nonzero
oscillator strengths are those with $s'=s+\alpha$.
The corresponding oscillator strengths can
be approximated as
\bea
   f^{(0)}_s(0) &=& \left(1-
      \frac{0.584}{1+2.64\,\gamma^{1.076}}\right)
         \,\frac{1+6\times10^{-6}\,\gamma}{
            1+0.247\gamma^{0.381}},
\label{f0lo}
\\
   f^{(1)}_{s,1}(0) &=& \left(1-
      \frac{0.584}{1+12\gamma^{1.43}}\right)
         \,\frac{1+9.8\times10^{-5}(s+1)\gamma}{
          1+1.585\gamma^{0.713}}
\nonumber\\&&\qquad\times\,
            [1+2s/(1+\gamma/30)]^{-1/4}.
\label{f0ri}
\eea
Equation (\ref{f0lo}) reproduces Eq.~(21) of \citet{P98},
but with a fixed typo, and \req{f0ri} additionally
generalizes it to nonzero $s$. 

For the moving atom and $\alpha=\pm1$, we keep only
transitions with $\Delta s=s'-s<4$, because the oscillator
strengths strongly decrease with increasing $\Delta s$.
First we define a field-dependent characteristic scale of $\Kp$,
\beq
K_1 = 42.9\,[1+\gamma/2.4+(\gamma/84)^2+(\gamma/380)^3]^{0.17}
  \mel e^2/\hbar.
\eeq
Then our approximation for the longitudinal polarization
reads
\bea
   f^{(0)}_s(\Kp) &=& f^{(0)}_s(0)\exp\left[-(c_1\Kp/K_1)^2\right]
        +
\frac{\exp[-(c_2\Kp/K_1)^{-c_3}]}{1+0.5\sqrt{K_1/\Kp}},
\nonumber\\&&
\eea
where
\[
   c_1 = \frac{4}{1+\gamma/1200},
\quad   c_2 = \frac{0.89}{1+\gamma/10^4},
\quad   c_3 = \max(1,10-\ln\gamma).
\]
For the main transition with circular polarization,
viz.{} $\alpha=\Delta s=1$, first
we introduce two functions describing
$f^{(1)}_{s,1}(\Kp)$ at small and large $\Kp$,
respectively,
\bea
   f^{(<)}_s(\Kp) &=& 1-p_1\,(\Kp/K_1)^2,
\quad
      p_1=(1+\gamma/800)^{-1/2},
\nonumber\\
   f^{(>)}_s(\Kp) &=& 2(s+1)\frac{\mel}{\mpr}
               \left[1-p_1\,(K_1/\Kp)^2\right],
\nonumber\eea
and truncate (replace by 0) the negative values of
these functions.
At any $\Kp$, our approximation reads
\beq
   f^{(1)}_{s,1}(\Kp) = f^{(<)}_s(\Kp)\,X(\Kp)
          + f^{(>)}_s(\Kp)\,\big[1-X(\Kp)\big],
\eeq
where
\beq
                 X(\Kp) = \frac{1}{1+(\Kp/K_1)^p},
\quad
   p = \frac{83}{1+[\ln(1+\gamma/90)]^2}.
\eeq
The oscillator strengths for the other considered
transitions are
\bea
   f^{(1)}_{s,\Delta s}(\Kp) &=& p_{s,\Delta s} (\Kp/K_1)^{2\Delta s -2}
             \max(0,1-q\Kp/K_1),
\\
         q&=&0.72+0.12\ln(1+\gamma/400),
\nonumber\\
   p_{s,2} &=& \frac{0.012\,[ 1+2s/(1+\gamma/30) ]^{3/4}}{
                1 + (\gamma/320)^{3/4}},
\nonumber\\
   p_{s,3} &=& 0.0016 + \frac{0.0055}{1 + (19/\gamma)^{1.8}
                  + (\Gamma/1270)^4},
\nonumber\\
  f^{(-1)}_{s,\Delta s}(\Kp) &=& 
          \frac{p'_{s,\Delta s}  (\Kp/K_1)^{2\Delta s+2} }{
            1+p''_{s,\Delta s} (\Kp/K_1)^{5\Delta s+5}},
\\
   p'_{s,1} &=& \frac{5\times10^{-5}}{1+110/\gamma}\,
        \left[1+\frac{2s}{1+\gamma/40}\right]^2,
\nonumber\\
    p'_{s,3} &=& 3.5\times10^{-4} \,\left[
         1+(230/\gamma)^{2/5} \right]^{5/2}
\nonumber\\
   p'_{s,2} &=& \frac{3.3\times10^{-5}}{1+(400/\gamma)^{1.2}}\,
        \left[1+\frac{2s}{1+\gamma/30}\right]^2,
\nonumber\\
   p''_{s,1} &=& \left[1+6s/(1+\gamma/40)\right]^{-1},
\nonumber\\
    p''_{s,2} &=& (p''_{s,1})^2,
\qquad
    p''_{s,3} = 1.
\nonumber\eea

% =============== JOURNAL ABBREVIATIONS =============
\newcommand{\artref}[4]{{#4}, {#1}, {#2}, #3}
\newcommand{\AandA}[3]{\artref{A\&A}{#1}{#2}{#3}}
\newcommand{\AIPC}[3]{\artref{AIP Conf.\ Proc.}{#1}{#2}{#3}}
\newcommand{\AnnPhysNY}[3]{\artref{Ann.\ Phys. (N.Y.)}{#1}{#2}{#3}}
\newcommand{\ApJ}[3]{\artref{ApJ}{#1}{#2}{#3}}
\newcommand{\ApJS}[3]{\artref{ApJS}{#1}{#2}{#3}}
\newcommand{\ApSS}[3]{\artref{Ap\&SS}{#1}{#2}{#3}}
\newcommand{\ARAA}[3]{\artref{ARA\&A}{#1}{#2}{#3}}
\newcommand{\JPB}[3]{\artref{J.\ Phys.\ B: At.\ Mol.\ Opt.\ Phys.}{#1}{#2}{#3}}
\newcommand{\jpb}[3]{\artref{J.\ Phys.\ B: At.\ Mol.\ Phys.}{#1}{#2}{#3}}
\newcommand{\MNRAS}[3]{\artref{MNRAS}{#1}{#2}{#3}}
\newcommand{\PL}[4]{\artref{Phys.\ Lett. #1}{#2}{#3}{#4}}
\newcommand{\PR}[4]{\artref{Phys.\ Rev. #1}{#2}{#3}{#4}}
\newcommand{\PRL}[3]{\artref{Phys.\ Rev.\ Lett.}{#1}{#2}{#3}}
\newcommand{\RMP}[3]{\artref{Rev.\ Mod.\ Phys.}{#1}{#2}{#3}}
\newcommand{\SSRv}[3]{\artref{Space Sci.\ Rev.}{#1}{#2}{#3}}
% ===================================================

\end{document}